\documentclass[11pt]{article}
\usepackage{a4wide}

\usepackage{type1cm}         

\usepackage{graphicx}        

\usepackage{newtxtext}       %
\usepackage{newtxmath}       

\usepackage{comment}

\usepackage[hidelinks]{hyperref}
\hypersetup{colorlinks=true,citecolor=blue,linkcolor=blue,
    filecolor=blue,urlcolor=blue}
\usepackage{amsmath}
\usepackage{mathdots}
\usepackage{amsfonts}
\usepackage{latexsym}
\usepackage{graphicx}
\usepackage{pgfplots}
\usepackage{todonotes}
\usepackage{bm}
\usepackage{svg}
\usepackage{color}
\usepackage{algpseudocode}
\usepackage{svg}

\pgfplotsset{compat=1.6}

\newcommand{\ignore}[1]{}

\newcommand{\bs}[1]{\mathbf{#1}}

\graphicspath{{./figs/}}

\begin{document}

\title{Approximations in Deep Learning}


\author{Etienne Dupuis$^1$, Silviu Filip$^2$, Olivier Sentieys$^2$, David Novo$^3$, Ian O'Connor$^1$, and Alberto Bosio$^1$\\
~\\
$^1$Univ Lyon, ECL, INL, Ecully, FR, firstname.lastname@ec-lyon.fr\\
$^2$Univ Rennes, Inria, IRISA, FR, firstname.lastname@inria.fr\\
$^3$LIRMM, Universit\'e de Montpellier, CNRS, FR, david.novo@lirmm.fr\\
}

\date{}

\maketitle

\abstract{The design and implementation of Deep Learning (DL) models is currently receiving a lot of attention from both industrials and academics. However, the computational workload associated with DL is often out of reach for low-power embedded devices and is still costly when run on datacenters. By relaxing the need for fully precise operations, Approximate Computing (AxC) substantially improves performance and energy efficiency. DL is extremely relevant in this context, since playing with the accuracy needed to do adequate computations will significantly enhance performance, while keeping the quality of results in a user-constrained range. This chapter will explore how AxC can improve the performance and energy efficiency of hardware accelerators in DL applications during inference and training.}

\tableofcontents

\section{Introduction}
\label{sec:introduction}

Deep Neural Networks (DNNs)~\cite{lecun2015deep}, and in particular, Convolutional Neural Networks (CNNs), are currently one of the most intensively and widely used predictive models in the field of machine learning. CNNs have been shown to give very good results for many complex tasks such as object recognition in images/videos, drug discovery, natural language processing, autonomous driving, and playing complex games~\cite{deng2013recent,krizhevsky2017imagenet,chen2015deepdriving,Silver_2016}.

Despite these benefits, the computational workload involved in CNNs is often out of reach for low-power embedded devices, and/or is still very costly when ran on datacenter-style Component-Off-The-Shelf (COTS) hardware platforms. To give an example, the amazing performance of AlphaGo~\cite{Silver_2016} required 4 to 6 weeks of training executed on 2000 CPUs and 250 GPUs for a total of about 600kW of power consumption (while the human brain of a Go player requires about 20W), which translates to over 2 TJ of energy consumption.
Thus, a lot of research effort from both industrials and academics has been concentrated on defining/designing custom hardware platforms supporting these types of algorithms, to improve performance and/or energy efficiency~\cite{wang2016dlau,chen2016eyeriss,liu2017throughput}.

CNNs show inherent resilience to insignificant errors due to their iterative nature and the underlying learning process. Therefore, an intrinsic tolerance to inexact computation is clear, and using the AxC paradigm to improve power and speed characteristics is, therefore, relevant~\cite{sung2015resiliency}. Indeed, CNNs mesh well with AxC techniques, especially with fixed-point arithmetic or low-precision floating-point implementations (it has been shown that even binary or ternary weights and arithmetic can be used), which moreover expose large fine-grain parallelism. They are therefore ideally suited for hardware acceleration using Field Programmable Gate Arrays (FPGAs) and/or Application-Specific Integrated Circuit (ASIC) implementations, as acknowledged by the large body of work on this topic. Although accelerators have demonstrated significant performance/energy gains compared to GPU/CPU implementations, they still require further efficiency to address future performance requirements~\cite{tann2017hardware}.

The goal of this chapter is to present an up-to-date view of state-of-the-art solutions applying AxC techniques to CNNs for both inference and training phases. It is structured as follows: Section~\ref{sec:Background} presents the background \& context of using DNNs, the main focus of the chapter. Section~\ref{sec:inference} overviews AxC methods found in the literature that improve deep neural network inference performance. Approximation techniques for improving the training part of neural network design, which accounts for the majority of computing time and resources, are presented in Section~\ref{sec:training}. Section~\ref{sec:accelerators} discusses DNN accelerator research and the dedicated approximation methods, whereas Section~\ref{sec:perspectives} presents incubent directions for AxC research in DL. Section~\ref{sec:conclusions} concludes the chapter.


\section{Background}
\label{sec:Background}
Artificial intelligence (AI) is a broad field of study focused on replicating or simulating the intelligence of living beings (human or not). It encompasses various methods and techniques. These range from design space exploration methods like ant colony optimization that focuses on finding increasingly efficient paths through simple random exploration and reward-based reinforcement, to more complex approaches such as genetic algorithms that evolve a population towards a hopefully optimized solution by iteratively picking the best candidates and mutating them. In the last couple of decades, Machine Learning (ML) algorithms have gained the most traction, producing effective predictions/answers based on some trained behavior/model. 




\subsection{Context: From AI to DNNs}


The ML subset of AI is focused on algorithms able to improve themselves through seeing already labeled input-output sample pairs and constructing models that attempt to match the expected outputs to this given data. An example is email filtering, deciding whether or not an email is spam based on its provenance, recipients, object, and other (meta)data. Generally, a model for this task is trained (\emph{i.e.,} it \emph{learns}) on a set of already labeled set of spam email data (the \emph{training data set}) until it reaches the desired behavior (the \emph{expected response}) with sufficient accuracy. It is then used with unseen data in the hope that it will still prove to be accurate (\emph{i.e.,}~generalize well). Evaluation of this generalization ability is frequently done on a so-called \emph{test} or \emph{validation} data set, different from the training data.

While email filtering can seem like a simple task, there are a plethora of use cases of varying difficulty where ML modeling is used, ranging from security (\emph{e.g.,}~in fraud detection) and business data analysis (\emph{e.g.,}~churn rate measurement) to computer vision, self-driving technologies, and other complex tasks. The model inputs can be both raw data or high-level features (for instance statistical aggregates of multiple input data samples) or other complex features that are task-dependent (\emph{e.g.,}~the presence of a horizontal line in an image). It is the task of the model to interpret this data and construct useful responses. Among the many tasks suitable for ML one can mention classification, regression, and semantic segmentation.


Traditionally, high-level features needed by a model were derived following a feature extraction step that was often performed by a human, requiring expert knowledge of relevant information. More recently, however, through the rise of DNNs in the ML ecosystem of approaches, this step can be performed automatically, the model is trained to discover relevant features, thus avoiding both the need for human expertise and the induced biases that might result from this.

\begin{figure}[ht]
\centering
\includegraphics[width=0.9\textwidth]{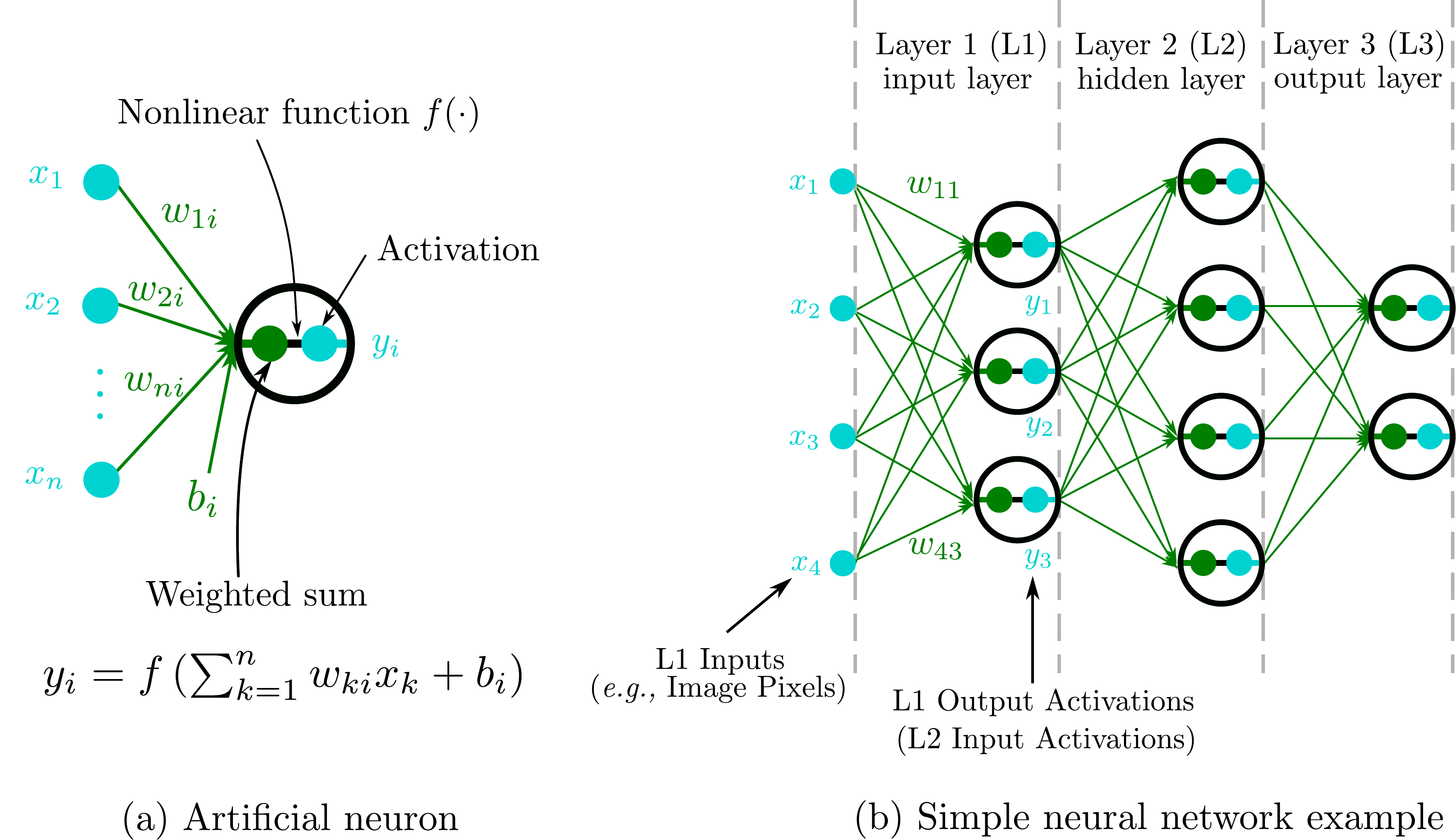}
\caption{A basic DNN example and the associated terminology (adapted from~\cite[Figure 1.3]{sze2020efficient}).}
\label{fig:basic_dnn}       
\end{figure}

Artificial neural networks are based on the notion that the computation performed by a neuron is centered around a weighted sum of its input values. This is shown in Figure~\ref{fig:basic_dnn}a, where multiple inputs $\{x_k\}_{k=1}^{n}$ are summed (scaled with \emph{weights} $\{w_{ki}\}_{k=1}^{n}$) together with an optional~\emph{bias} term $b_i$. The neuron \emph{output} $y_i$ is determined by the application of a~\emph{nonlinear activation function} $f$ to this weighted sum. There are many activation functions used in practice, but among the most common are $f(x) = \text{ReLU}(x) := \max\{0,x\}$ and $f(x) = \tanh(x) := (e^{x} - e^{-x})/(e^{x}+e^{-x})$. 

Such neurons are grouped together to form \emph{layers}. The present chapter is focused on feedforward networks, where the outputs of a layer are then used as inputs for subsequent layers\footnote{There are classes of \emph{recurrent} neural networks that allow outputs of a layer to be connected to inputs of previous layers. While they are not discussed here any further, they are frequently used to process sequential data (\emph{e.g.,}~speech, text).}. This is exemplified in Figure~\ref{fig:basic_dnn}b. The inputs and outputs of a layer are also known as~\emph{input} and~\emph{output activations}, respectively. When discussing visual data, they are also known as input and output~\emph{feature maps}. The first and last layer in the network are generally known as the~\emph{input} and~\emph{output layer}, respectively. In between them, there is a number of intermediate layers, called~\emph{hidden layers}. The main characteristic of DL and DNNs is that the number of hidden layers can grow quite large, from two layers up to even one thousand.

\begin{figure}[ht]
\centering
\includegraphics[width=0.9\textwidth]{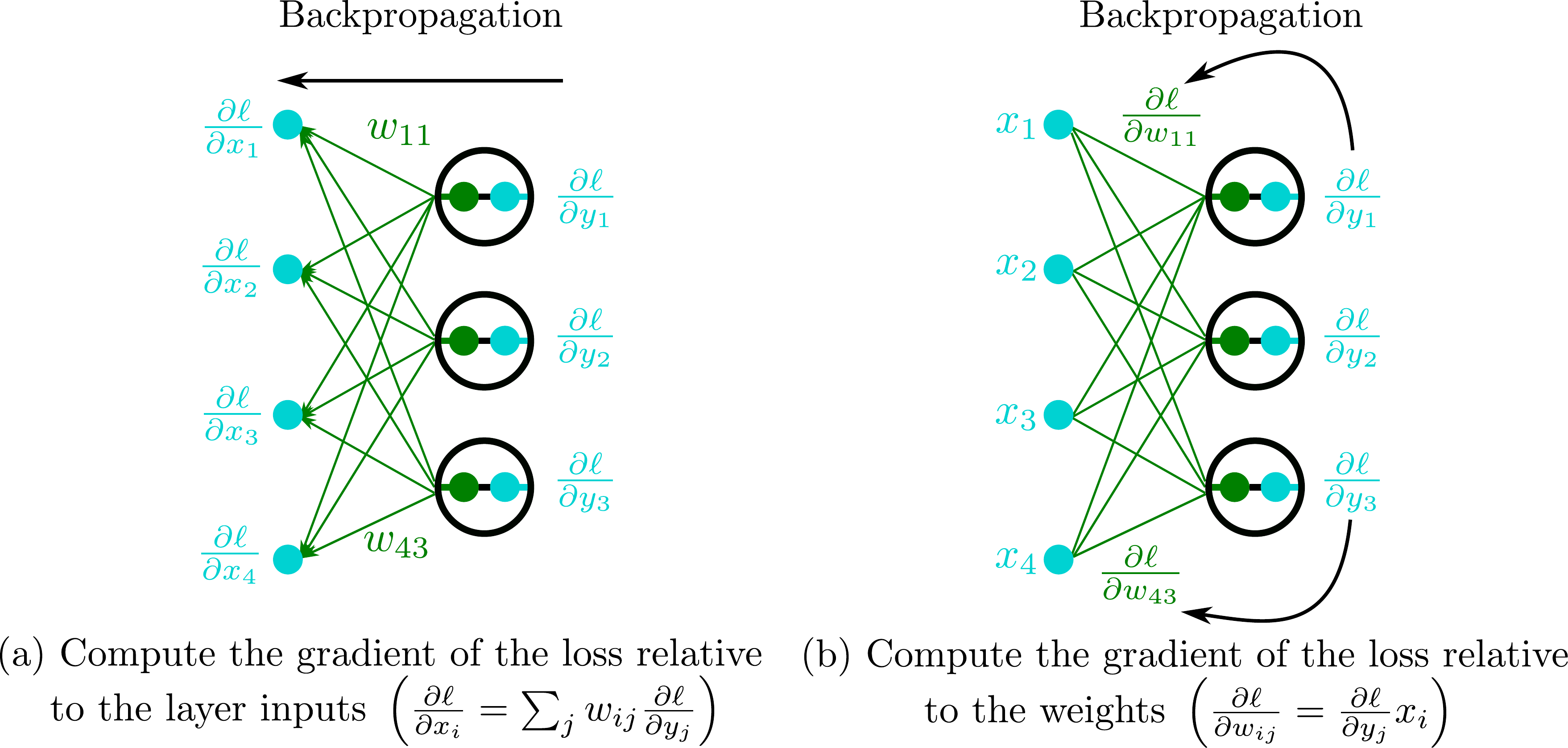}
\caption{A backpropagation example through a neural network (adapted from~\cite[Figure 1.6]{sze2020efficient}).}
\label{fig:backprop}       
\end{figure}

The process of using an artificial neural network with a set of given parameters (\emph{e.g.,} weights and bias terms) is called~\emph{inference}. For the neural network to be useful, its inference output has to match as closely as possible an expected/ideal output. This is measured through a~\emph{loss} function $\ell$ that compares how far the resulting output on (subsets of) the training and test data sets is to the expected output. Thus, the goal of \emph{training} a neural network is to find/learn a set of parameters that minimizes the average loss over a large training set.

To train a network, its weights ($w_{ij}$) are usually updated using a form of Stochastic Gradient Descent (SGD) iterative optimization process. This means that weight is updated by a scaled version of the partial derivative of the loss function $\ell$ with respect to the weight. In the most basic form, at iteration $t$, the weight update formula is given by:
\begin{equation}\label{eq:sgd}
    w_{ij}^{t} = w_{ij}^{t-1}-\alpha\dfrac{\partial \ell}{\partial w_{ij}^{t-1}},
\end{equation}
where $\alpha$ is called the learning rate\footnote{The deep learning optimization literature describes many ways how to perform the parameter updates and how to choose the learning rate.}. The partial derivatives of $\ell$
can be computed efficiently through a process called~\emph{backpropagation}~\cite{rumelhart1986learning}. It is effectively an application of the \emph{chain rule} from calculus, and it works by passing values backward through the network to compute how $\ell$ is affected by each weight. At each layer, the procedure is twofold and is exemplified in Figure~\ref{fig:backprop}. To backpropagate through a layer: (a) compute the gradient of the loss with respect to the weights, $\partial \ell / \partial w_{ij}$, from the layer inputs (\emph{i.e.,}~the forward activations $x_i$) and the gradients of the loss relative to the layer outputs, $\partial \ell / \partial y_j$; and (b) compute the gradient of the loss relative to the layer inputs, $\partial \ell / \partial x_i$, from the layer weights, $w_{ij}$, and the gradients of the loss relative to the layer outputs, $\partial \ell / \partial y_j$. 

Computing the gradients of the loss function $\ell$ over the entire dataset is generally much too complicated in practice, which is why the loss is usually taken only on a (small) subset, called a~\emph{mini batch}, of the training data. The use of batches allows taking advantage of single instruction multiple data (SIMD)-like parallelism on modern GPUs while keeping the complexity of gradient computation manageable. A complete iteration of the training process is called an~\emph{epoch} and requires passing through all of the mini-batches, applying~\eqref{eq:sgd} for each one of the corresponding average losses $\ell$. Training is carried out for several epochs until convergence to an appropriate solution is reached.

Both inference and training amount in most part to the same type of computations (\emph{i.e.,}~matrix/vector additions and multiplications). There are important differences, however. For one, as the previous paragraph suggests, training is much more expensive, since apart from passing through the entire training data multiple times, it also requires that intermediate outputs and partial derivatives be stored
when performing backpropagation. Secondly, due to the gradient update rule, the precision requirements for training are generally higher than for inference, thus also affecting performance. The effect is that the inference quantization techniques that will be discussed in this chapter are not usually directly applicable to training as well.

\subsection{Deep Learning Landscape}

While artificial neural networks have a long history dating as far back as the 1940s, practical applications using digital neurons did not arrive until the late 1980s, when the LeNet-5~\cite{le1989handwritten, LeNet-5} network architecture was used for hand-written digit recognition. It is only in the early 2010s however, with the synergy of three major factors, that artificial neural network models have started to take off, under the names~\emph{deep learning} and~\emph{deep neural networks}. These factors are: (1) the availability of large and labeled datasets that are needed to train complex models; (2) the advance in computational power of units such as GPUs that allow DNN training to be executed in reasonable time (days or weeks instead of years); (3) development of new algorithmic techniques (\emph{e.g.,} the Adam gradient descent optimization algorithm~\cite{kingma2014adam}) that enable improved accuracy at a larger scale.

The importance of large and comprehensive datasets cannot be overstated. If not careful, a small training dataset used in conjunction with a complex DNN can easily lead to \emph{overfitting} (\emph{i.e.,}~the model matches the training data extremely well but does not generalize to unseen data accurately).  For computer vision, arguably the most popular dataset in recent years has been ImageNet~\cite{imagenet}, a collection of one million high-resolution images that are generally associated with the ILSVRC~\cite{ILSVRC15} image recognition contest that uses 1000 labeled categories. Smaller datasets such as MNIST~\cite{mnist} and CIFAR~\cite{KrizhevskyCIFAR} have also been used extensively in DNN research for inference and training acceleration.




Apart from the data, the choice of model (network architecture and associated parameters) is also crucial in the success of a DL approach. In what follows (Section~\ref{par:cnn}), our focus is on CNN models suited to process visual data.


The current surge of interest in DL is also facilitated by the availability of tools and frameworks that allow for the easy prototyping and design of DNN models. Prominent examples include Tensorflow~\cite{tensorflow} and Pytorch~\cite{pytorch}. The open-source nature of these alternatives offers the possibility to design extensions that can be leveraged throughout a model's lifecycle (from initial prototype to deployment).


Depending on the intended use of DNN models, they can be found in different environments with various computing power and energy consumption characteristics. At one end of the spectrum there are edge devices characterized by low-power and limited computational capabilities, while at the other end power-hungry cloud devices with a high-performance computing profile are dominant. 

\subsection{Convolutional Neural Networks}
\label{par:cnn}


The most basic layer in a feedforward network is the Fully Connected (FC) or dense layer. It is characterized by the fact that each neuron in the layer is connected to all the neurons in the previous layer. FC layers are parameterized by the number of neurons they contain. An example is shown in Figure~\ref{fig:basic_dnn}, which only has FC layers.

 
 \begin{figure}[ht]
\centering
\includegraphics[width=0.9\textwidth]{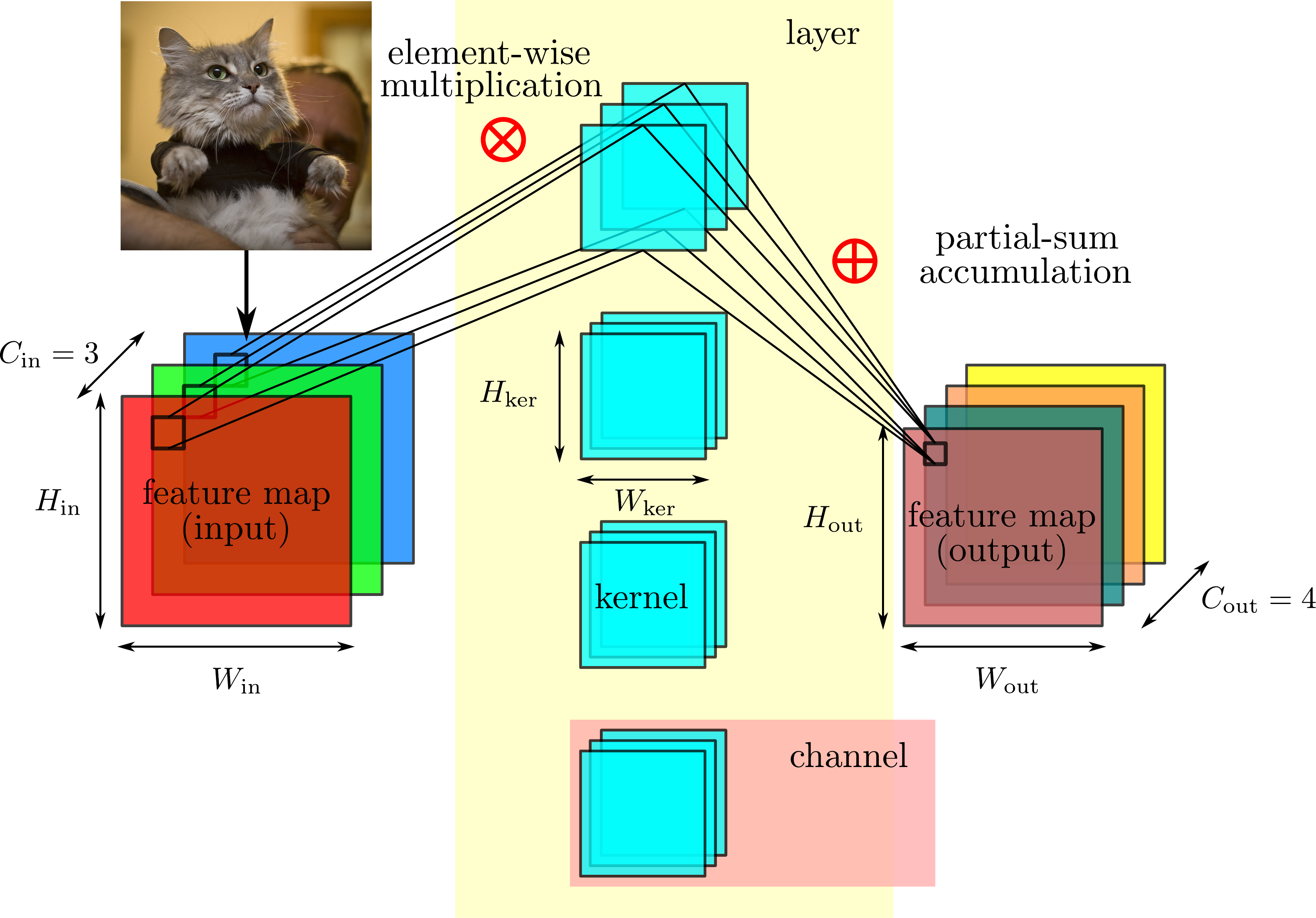}
\caption{Expanded view of a typical 2D CONV layer inside a CNN.}
\label{fig:convlayer}
\end{figure}
 
 While the expressive power of networks using only FC layers is impressive, it comes at the cost of a very large number of connections (and hence network parameters), making them hard to train and easily prone to overfitting. This is why other types of structured layers, with fewer parameters, but which are more efficient for certain tasks, have been explored. In the case of visual data, this has led to the development of CNNs, a staple of DL today.
 
 The main elements that have led to the introduction of CNNs are Convolutional (CONV) layers, composed of high-dimensional convolutions that allow extraction of shift-invariant features from the input. An example is Figure~\ref{fig:convlayer}, showing a traditional 2D CONV layer. In this context, the input activation is structured as a 3D set of input feature maps, with input width ($W_\text{in}$), input height ($H_\text{in}$) and input channel ($C_\text{in}$) dimensions. The weights of the layer are structured as a 3D filter, with kernel width ($W_\text{ker}$), kernel height ($H_\text{ker}$) and input channel ($C_\text{in}$) dimensions. For each input channel, the corresponding input feature map is transformed through a 2D convolution with the appropriate kernel in the filter. The convolution results at each point are summed across all the input channels to generate the output partial sums. The results of these partial sums comprise one output feature map with output width ($W_\text{out}$) and output height ($H_\text{out}$) dimensions. Several 2D filters can be stacked together to generate additional output channels, denoted with $C_\text{out}$ in this case.

\begin{figure}[ht]
\includegraphics[width=\textwidth]{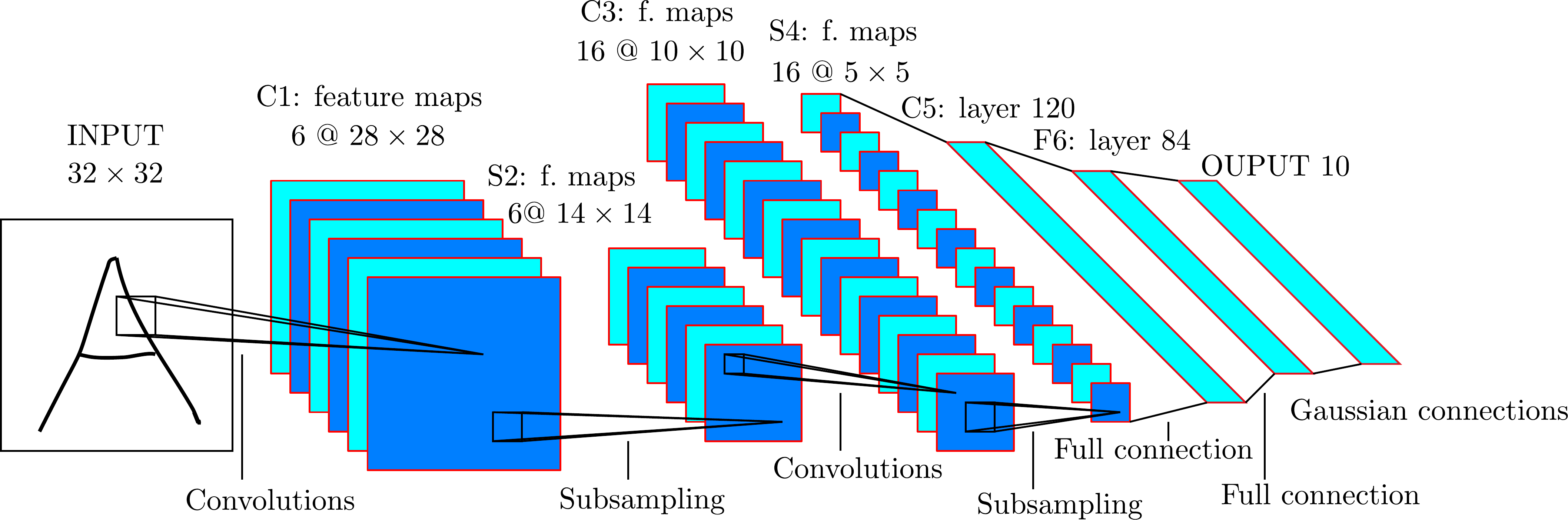}
\caption{A visual representation of LeNet-5 (adapted from~\cite[Fig.~ 2]{LeNet-5}), an early example of a CNN that promoted the subsequent development of Deep Learning. It contains the main layers that are usually found in CNNs: convolutional, pooling and fully connected.}
\label{fig:lenet_5}       
\end{figure}

Depending on the size of 2D kernels and their count, the output feature maps can be large and deep, motivating the use of pooling (\emph{i.e.,}~subsampling) layers that reduce the scale of feature maps. Pooling is similar to convolution,  with a kernel sliding over the input matrix, but instead of performing matrix multiplication, an aggregation operation function is applied. The most common such operations are taking the maximum element or the average. A visual example of a simple CNN mixing in all these layers is given in Figure~\ref{fig:lenet_5}.


Another frequently used layer is Batch Normalization~\cite{ioffe2015batch} (BN). It contains two trainable parameters that are used to re-center and re-scale the distribution of the values of a feature map, to improve training performance. While there are also more recent and complicated layers, such as depth-wise convolutions~\cite{Howard2017Mobilenet} or Inception modules~\cite{GoogLeNetInception}, their specifics are not important for the rest of this chapter.


\subsection{Performance and Energy Profiles of Recent Models}

To gauge the complexity of current DNN models, there is a need for a set of metrics that allow for a fair comparison between models. In this study, the metrics used are (1) the model accuracy over a validation dataset, (2) the total number of weights in the model, and (3) the number of FLOating-Point operations (FLOPs) necessary to carry out one complete inference. Accuracy is measured in terms of the frequently used top-1 and top-5 percentages (\emph{i.e.,}~the proportion of correct predictions on the labeled validation dataset and the probability that the correct result is among the top five predictions). The number of weights allows estimating the total memory storage requirements for the model, whereas the FLOP count hints at the required computing power needed to execute the model at a certain frequency.
\begin{table}[ht]
\scriptsize
\centering
\begin{tabular}{l|lll|ll}
\textbf{Model Name}             & AlexNet~\cite{alex_krizhevsky_imagenet_1905}         & GoogLeNet~\cite{GoogLeNetInception}       & ResNet-50~\cite{ResNet}        & MobileNet V2~\cite{s2018mobilenetv2}    & EfficientNet B1~\cite{Tan2019Efficientnet} \\
\textbf{Year}                   & 2012            & 2014            & 2016             & 2018            & 2019            \\
\textbf{Top-1 accuracy}         & 57.2\%          & 69.8\%          & 76.2\%           & 72.0\%          & 79.1\%  \\
\textbf{Top-5 accuracy}         & 84.7\%          & 93.3\%          & 92.97\%          & 90.6\%          & 94.4\% \\
\textbf{Number of Weights}       & 62M             & 6.4M            & 26M              & 3.5M            & 7.8M            \\
\textbf{FLOPs}                  & 1.5B            & 2B              & 4.1B             & 0.3B            & 0.7B       

\end{tabular}
\caption{Recent evolution of DNNs for image classification on the ImageNet dataset.}
\label{table:DNNcomparison}
\end{table}


Table~\ref{table:DNNcomparison} shows a comparison using these metrics on some popular DNNs for image classification on the ImageNet dataset (adapted from~\cite{PapersWithCodeImagenet}). For a long time, the only metric of interest was the network accuracy, resulting in models that were costly to train and operate. The cost of training and inference became so large at one point that there is now an open engineering consortium called MLCommons\footnote{\href{https://mlcommons.org/en/}{https://mlcommons.org/en/}} that benchmarks DL models and fosters innovation in the field. Thus, there is an increasing interest for faster, lighter, and overall more efficient models that are compatible with edge device resource constraints and operate more efficiently in the cloud. The last two columns in Table~\ref{table:DNNcomparison} reflect this, with newer network models achieving competitive accuracy with less memory and a smaller FLOP count.

Some examples of the scale at which modern DNN training costs stand for recent NLP models are given in Table~\ref{table:DNNTrainingCost} (adapted from~\cite[Table 3]{strubell2019energy}) and showcase the significant resources needed for training state-of-the-art models.
\begin{table}[ht]
\centering
\begin{tabular}{lcccc}
\textbf{Model}             & \textbf{Hardware}         & \textbf{Power (W)}       & \textbf{Hours}        & \textbf{CO$_2$e (lbs)} \\
\hline\hline
Transformer$_\text{base}$~\cite{vaswani2017attention} & P100$\times$8 & $1415.78$ & $12$ & $26$ \\
Transformer$_\text{big}$~\cite{vaswani2017attention} & P100$\times$8 & $1515.43$ & $84$ & $192$ \\
ELMo~\cite{peters2018deep} & P100$\times$3 & $517.66$ & $336$ & $262$ \\
BERT$_\text{base}$~\cite{devlin2018bert} & V100$\times$64 & $12041.51$ & $79$ & $1438$ \\
BERT$_\text{base}$~\cite{devlin2018bert} & TPUv2$\times$64 & --- & $96$ & --- \\
NAS~\cite{so2019evolved} & P100$\times$8 & $1515.43$ & $274.12$ & $626.155$ \\
NAS~\cite{so2019evolved} & TPUv2$\times$1 & --- & $32.623$ & --- \\
GTP-2~\cite{radford2019language} & TPUv2$\times$32 & --- & $168$ & --- \\
\end{tabular}
\caption{Estimated cost of training recent NLP models in terms of power, time and CO$_2$ emissions.}
\label{table:DNNTrainingCost}
\end{table}


The need for efficient DL computations coupled with the resilience of DNNs to approximation (due to the stochastic nature of training methods and a high level of inner redundancy~\cite{Chippa2013AnalysisAC}) has paved the way for the development of a large number of approximation methods, a part of which are described in the rest of this chapter.




\section{Approximation for Inference}
\label{sec:inference}

DNN inference is a very computation-intensive task, having large memory and computation power requirements. For example, inference on a single image using the original ResNet34~\cite{he2016deep} model requires 3.6 billion FLOPs and storing 22 million weights plus temporary feature maps. While the execution of such tasks has moved from traditional CPUs having a latency-oriented design to more parallel hardware like GPUs or even custom ASICs / FPGAs, inference is still a costly task, and thus susceptible to benefit from performance improvements when using approximate computing. Consequently, this section describes AxC methods found in the literature that improve deep neural network inference performance.

\begin{figure}[t]
\centering
\includegraphics[width=0.6\textwidth]{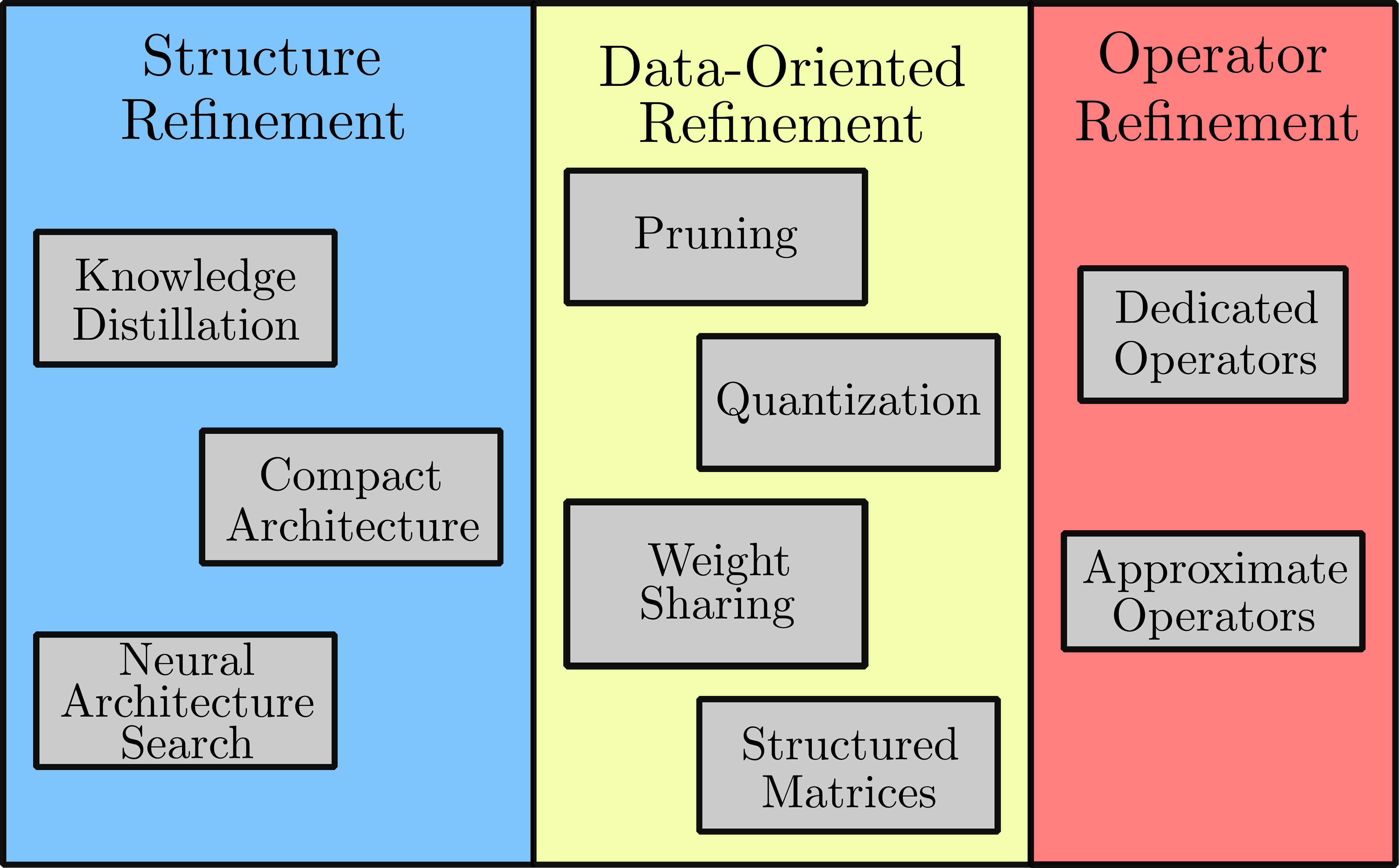}
\caption{Different types of approximation techniques for DNN inference.}
\label{fig:approx_inference}       
\end{figure}

One can distinguish three different classes of methods (see Figure~\ref{fig:approx_inference}), usable in isolation or combined, to approximate DNN inference. The first one,~\emph{structure refinement transformations}, includes the methods that modify the computational structure (\emph{i.e.,}~the network layers and their parameters) of the input model. Some notable examples include knowledge distillation~\cite{hinton2015distilling,tang2020understanding} which uses the model as a teacher to help train smaller students models or compact architectures~\cite{Howard2017Mobilenet, squeezenet} where layers are transformed into more hardware friendly ones. The second class,~\emph{data-oriented refinement transformations}, focuses on optimizing the finite precision data representation(s) of the model while maintaining the initial computational structure intact. Notable examples are~\emph{pruning}~\cite{cun1990optimalbrain, Frankle2018LotteryTickets} (\emph{i.e.,}~setting less important parameters to zero to increase sparsity) and~\emph{quantization}~\cite{zhou2016dorefa, zhou2017incremental} (\emph{i.e.,}~changing the types of the parameters and intermediate results to more efficient representations). While network structure refinement substantially changes the network structure, data type refinement does not, giving the possibility to emulate the inference of the approximated network in the original structure to measure its accuracy loss.

The third class of approaches relies on ~\emph{operator refinement transformations} which modify the arithmetic operators used inside the CNN implementation (\emph{e.g.,}~addition and multiplication) to further improve energy efficiency. Such methods are not discussed any further here since they mainly depend on the hardware implementation of the CNN. More details on the approximate operators that fall in this class can be found in Chapters 3 and 4. 

In the rest of this section, data-oriented refinement methods are covered. 
Specifically, Section~\ref{subsec:quantization} gives an overview of various quantization methods, Section~\ref{subsec:weightsharing} discusses weight sharing approaches, whereas pruning is analyzed in Section~\ref{subsec:pruning}.

\subsection{Quantization}
\label{subsec:quantization}

Full precision DNNs usually rely on 32-bit floating-point values for representing parameters. For standard backpropagation-based training, using high precision weights makes sense since the gradient update rule generally modifies these weights by a small factor of the corresponding gradient terms. While full precision~\texttt{float32} DNNs offer excellent result quality, they can generally be compressed and accelerated using lower precision arithmetic with minimal or no loss in the accuracy. Methods for addressing data quantization in DNNs are varied, ranging from simple binary and ternary networks to larger fixed-point and custom floating-point formats. This section gives an overview of the main ones.

Analysis of existing approaches relies on various aspects, such as (1) what parts of the network are being quantized, (2) homogeneity/heterogeneity of the number formats used inside the layers, (3) the type of representations being used, and (4) how and when is quantization performed (during or after the network has been trained).

    \textbf{What to quantize.} The most obvious quantization targets are the~\emph{network parameters} (\emph{e.g.,}~weights and biases). Reducing the number of bits used to represent them primarily brings a memory footprint reduction for on-device storage of the network. Latency improvements are potentially achievable with binary, ternary and bit-shift (\emph{i.e.,} power of two values) quantized parameters~\cite{courbariaux2016binarizeda,li2016ternary,rastegari2016xnor}. More generally, if faster execution times are to be obtained, \emph{activation function inputs and outputs} also need to be quantized. An example is~\cite{jacob2018quantization}, which proposes an efficient 8-bit integer quantization scheme for both weights and activations. Additionally, one can quantize the weight and activation~\emph{gradients} used during backpropagation (see for instance~\cite{zhou2016dorefa, wu2018training}) to accelerate training, an aspect discussed in Section~\ref{sec:training}.

    \textbf{When and how to perform quantization.}
    There are two established ways quantization can be performed for efficient inference and a third, emerging method. 
    
    The first among the established approaches is~\emph{Quantization-Aware Training} (QAT). The idea is to use a network parameter update procedure for several epochs (starting from scratch or after a baseline \texttt{float32} training method is run) to adjust parameters in the quantization format(s) such that generalization accuracy is hopefully kept the same or is at worst minimally degraded. Much research has focused on such fine-tuning methods (see for instance~\cite{rastegari2016xnor,zhou2016dorefa,zhou2017incremental,choi2018pact,jacob2018quantization,zhang2018lq}), mainly because they achieve good results, especially for extremely low precision formats (\emph{i.e.,}~binary and ternary encodings).
    
    While training is a powerful approach to compensate for a model's accuracy drop due to quantization, it is not always applicable in real-world scenarios (\emph{e.g.,} for online learning) since it is costly, time-consuming and generally requires a full-size training dataset. This can be a problem when the data is proprietary, privacy and regulatory issues are in effect (\emph{e.g.,} medical data that cannot be uploaded to the cloud for remote processing), or when using pre-trained off-the-shelf models for which data is no longer available. As such, there has been a push for faster~\emph{Tost-Training Quantization} (PTQ) methods without any fine-tuning. It has been observed that for down to 8-bit word lengths, PTQ results are close to full precision ones for several models~\cite{banner2019post} (\emph{e.g.,} AlexNet, VGG, and ResNet), but it becomes significantly more difficult to maintain accuracy when targeting lower precision formats. Work focused on PQT includes~\cite{banner2019post,cai2020zeroq,choukroun2019low,nagel2019data,zhao2019improving}.
    
    A possible issue with QAT and PQT methods is that both generate networks that are \emph{sensitive} to how quantization is carried out (\emph{e.g.,} the target word length). As such, there has been recent work~\cite{alizadeh2020gradient,shkolnik2020robust} on methods for~\emph{robust quantization} that provide intrinsic tolerance of the model to a large family of quantization formats and policies by directly specifying it in the training loss function. Such approaches are interesting for battery-powered edge devices, where depending on the state of charge, a network model capable of operating effectively at various quantization levels would be highly beneficial.
        
    \textbf{Granularity of applying a quantization format.} Initially, quantization approaches were homogeneous, with one word length being used for the entire network. This is the case for early works on binary~\cite{courbariaux2015binaryconnect} and ternary~\cite{li2016ternary} weight networks, for instance. Such approaches can suffer from significant accuracy loss since different layers tend to have different sensitivities to quantization levels/noise. Subsequent work has focused more on a heterogeneous, layer-wise optimization of the quantization format~\cite{zhou2017adaptive,wu2018mixed,wang2019haq,dong2019hawq,dong2019hawqv2,cai2020zeroq}.
    
    There have been various metrics proposed to estimate the overall effect of a fixed-point quantization format inside a layer on the overall accuracy of the network. One example is~\cite{lin2016fixed}, which uses a Signal to Quantization Noise Ratio (SQNR) to empirically measure how suitable a fixed-point format is. The approach in~\cite{zhou2017adaptive} generalizes the work from~\cite{lin2016fixed} using an adversarial noise to formulate the quantization error. Another adaptive quantization method is~\cite{khoram2018adaptive}, which uses the loss function gradient to determine an error margin for each parameter such as to not degrade accuracy and assign a precision accordingly. Recent work~\cite{dong2019hawq,dong2019hawqv2,shen2020q} also proposes using second-order information (Hessian-based) to gauge the sensitivity of each layer. From an Information Theory perspective,~\cite{zhu2018adaptive} uses the entropy of weights and activations as a saliency indicator to set fixed-point quantization levels at each layer. Another popular statistical sensitivity measure is based on the Kullback-Leibler divergence, which is used to measure layer sensitivity in~\cite{wang2019haq,cai2020zeroq} and is a core component for fine-tuning low precision integer weights in NVIDIA's TensorRT inference acceleration library.
    
    On a different granularity level,~\cite{park2018value} proposes looking at the distribution of weight values over the entire network to aggressively quantize weights in dense regions and more gently those in sparse ones. Compared to \texttt{float32} baselines, such an approach can achieve under 1\% accuracy loss for large networks (ResNet-152 \& DenseNet-101) with a 4-bit format in the dense areas and a 16 bit one for the sparse regions ($<1\%$ of parameters).

    \textbf{Quantization formats.} There have been various representations used to quantize deep neural networks. At the extreme, there are~\emph{Binary Neural Networks} (BNNs), where weights and activations are stored with one of two possible values. If a $\{0, 1\}$ (or equivalently a $\{-1, +1\}$) encoding is used, then multiplications can be implemented efficiently using XNOR gates, making BNNs compelling on FPGA and ASIC targets, but also for emerging computing paradigms such as neuromorphic~\cite{esser2016cover} or in-memory computing~\cite{sun2018computing}. 
        
    Among the first investigations of binary networks is BinaryConnect~\cite{courbariaux2015binaryconnect}, which maintains a full precision copy of the weights to be updated during backpropagation, but are binarized for inference. Activations are kept in full precision, meaning full precision accumulations are still required during the forward propagation. The effect of binary activations is considered in~\cite{rastegari2016xnor,courbariaux2016binarizeda,courbariaux2016binarizedb}. These early papers are the basis for most subsequent research on BNNs.
        
    The XNOR-Net approach~\cite{rastegari2016xnor} expands on the initial BNN ideas by proposing a model where a gain term is added to the network at the level of each dot product in the convolutional layers. Computed from statistics of weights and activations before binarization, the gain was a way to improve the accuracy of BNNs on the ImageNet dataset. Such gain terms are nevertheless costly to compute in practice, which is why later work modified their use. For instance,~\cite{zhou2016dorefa} proposes gain terms that are only based on the non-binarized weights of the network, meaning that they never need to be recomputed after training. Additionally,~\cite{tang2017train} also advocates binarizing fully connected layers by adding neuron-specific scaling factors, further improving compression without a drastic decrease in the accuracy. A generalization of the BNN concept to multiple binary bases used for quantizing weights and activations is presented in~\cite{lin2017towards}, further reducing the accuracy gap between full precision and binary architectures, at the expense of a higher computational cost (compared to previous BNN methods). Changes to the backpropagation process in BNN training~\cite{darabi2018bnn+} can also be effective for limiting accuracy loss.
        
    \emph{Ternary neural networks} offer a better representation of the (pseudo) normal distribution of weights that is frequently observed after training. For instance,~\cite{hwang2014fixed} achieved good results on small networks with weights quantized to $\{-1, 0, +1\}$ and 3-bit fixed-point activations. For greater flexibility,~\cite{li2016ternary} proposes using a threshold $\alpha$ for picking the ternary weights ($-1$ if $w<-\alpha$, $0$ if $|w|<\alpha$ and $+1$ if $w\geqslant \alpha$), while keeping activations in full precision. This is further expanded in~\cite{zhu2016trained}, which uses ternary weights from a set $\{-\alpha^n, 0, +\alpha^p\}$, where $\alpha^n$ and $\alpha^p$ are learnable parameters. By also quantizing activations to 8-bits and adding residual edges to branches in the architecture that are sensitive to quantization,~\cite{kundu2017ternary} offers comparable accuracy results to~\texttt{float32} for a ResNet-101 model on the ImageNet dataset, with no additional low-precision (re)training. In a more aggressive compression strategy,~\cite{wan2018tbn} proposes the use of ternary activations $\{-1, 0, +1\}$ and binary scalable weights $\{-\alpha, +\alpha\}$.

    Extremely low-bit width networks like the ones just presented are susceptible to non-negligible accuracy loss, which is why there has been work focusing on non-binary~\emph{integer} and~\emph{fixed-point}-based quantization. Among the early proponents of integer quantization, there is~\cite{zhou2016dorefa}, which extends the idea of BNNs to arbitrary word lengths for weights, activations, and gradients. For fixed-point arithmetic,~\cite{lin2016fixed} explored the use of various bit width combinations ($4, 8$ and $16$ bits) of weights and activations. Notable results with integer arithmetic are presented in~\cite{jacob2018quantization}, which showcases how 8-bit integer quantization on ARM CPUs can achieve near-identical accuracy compared to baseline \texttt{float32} models based on MobileNet architectures for classification and detection tasks, but with improved on-device latency. Good quantization results with 4-bit weights and activations are presented in~\cite{banner2019post} by combining three complementary methods for minimizing quantization error at the tensor level. Heterogeneous/mixed-precision quantization approaches also heavily focus on integer/fixed-point formats~\cite{wu2018mixed,wang2019haq,dong2019hawq,dong2019hawqv2,cai2020zeroq}.
    
    One problem with low precision integer/fixed-point formats is that they have limited dynamic range, which might make them inappropriate, especially for networks used in Natural Language Processing (NLP) tasks, where weights tend to have values that are more than $10\times$ larger than the largest magnitude values found in popular CNNs~\cite[Fig.~1]{tambe2020algorithm}. While not that widespread, there has been some work looking into low precision \emph{floating-point} quantization for CNN inference. For instance,~\cite{settle2018quantizing} explores the use of up to 8-bit (scaled) floating-point formats for weight and activation quantization in classification networks such as GoogLeNet, ResNet, and MobileNet, without any accuracy degradation. More recently,~\cite{wu2020low,wu2020phoenix} show how an 8-bit floating-point quantization format (4-bit mantissa and 3-bit exponent) can be used in FPGA-based accelerators for deep CNN inference, without any retraining. Another approach~\cite{tambe2020algorithm} consists of an \emph{adaptive} floating-point quantization method, where the exponent range of quantized values is dynamically shifted at each network layer (through changing the bias term of the exponent), yielding competitive results on NLP networks and tasks. 
        
    At a coarser level, it is also possible to improve dynamic range by sharing the exponent between parameters, storing only the mantissa and one copy of the exponent. This is the so-called \emph{Block Floating-Point} (BFP) format. For instance,~\cite{song2017computation} propose using BFP with an 8-bit mantissa for weight storage, showing negligible to no accuracy loss on CNN workloads (VGG16, ResNet-18, ResNet-50, and GoogLeNet-based networks). On the FPGA side of things,~\cite{lian2019high} showcases a BFP-based CNN accelerator design that uses 16-bit activations and 8-bit weights, reducing memory requirements compared to a \texttt{float32} baseline without any retraining/fine-tuning. Another way to increase the dynamic range is to employ a~\emph{logarithmic representation}, which also allows multiplications to be replaced with simple binary shift operations. For instance,~\cite{miyashita2016convolutional} shows that a $\log$ representation can achieve higher classification accuracy than fixed-point formats operating at the same word length. 8-bit log floating-point quantization was also shown~\cite{johnson_rethinking_2018} to perform close to baseline \texttt{float32} values with several CNN classification networks.
    
\begin{figure}[ht]
\centering
\includegraphics[width=0.7\textwidth]{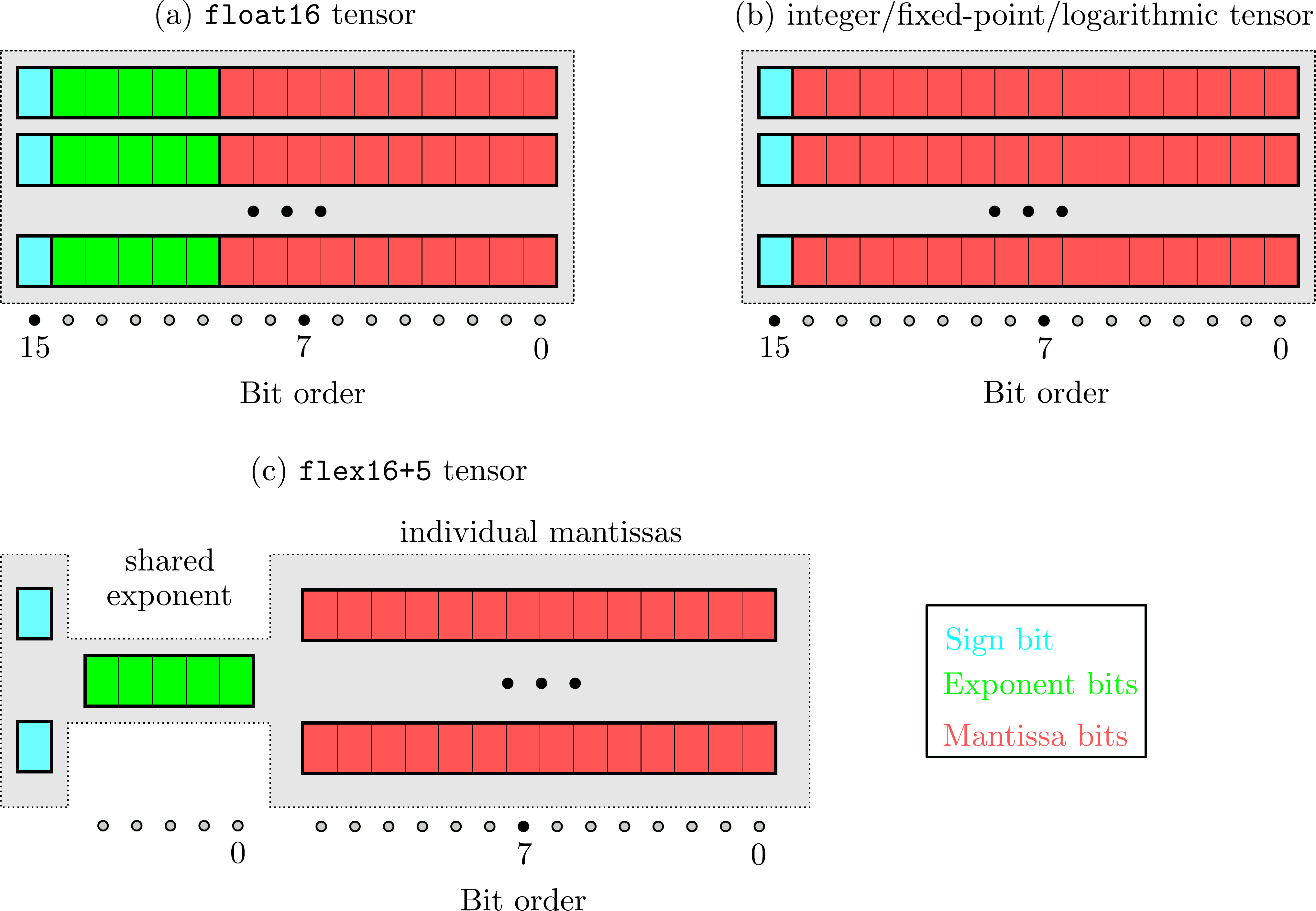}
\caption{Diagrams for bit representations of various numerical formats discussed in the context of DL quantization in this chapter. Red, green and blue shading are used to represent mantissa ($M$), exponent ($E$), and sign ($S$) bits respectively. In (a), the 16-bit IEEE 754 \texttt{float16} \emph{floating-point} format is shown (corresponding to $(-1)^S\times 2^{E-15}\times 1.M_2$ for normalized values), with 1 sign bit, 5 exponent bits and 10 mantissa bits. (b) illustrates a 16-bit \emph{signed integer} format. By choosing a~\emph{fixed} splitting point for integer ($I$) and fractional ($F$) parts in the mantissa ($M:=I.F$), it can also serve as a representation for a \emph{fixed-point} format (namely to $(-1)^S\times I_2.F_2$). Additionally, (b) can represent a form of \emph{logarithmic number system} (see for instance~\cite{fu2010fpga}), with the encoded value being $(-1)^S\times 2^M = (-1)^S\times 2^{I.F}$. Part (c) exemplifies a \emph{block floating-point} format, namely the \texttt{flex16+5} format~\cite{koster2017flexpoint} with a 15-bit mantissa and 5-bit shared exponent.}
\label{fig:nb_formats}       
\end{figure}
    
    A summary of these aforementioned formats (minus the binary and ternary encoding that generally require just 1 or 2 bits to represent) is given in Figure~\ref{fig:nb_formats}.
    
    Looking at the~\emph{value distribution} of the data (weights and activations) is a good way to explore what number formats and/or encodings are better suited for a particular network model. Uniform precision was the go-to alternative for a long time, but more recent work is concentrated around non-uniform quantization. This is because the actual distributions of trained weights tend to follow bell-shaped curves. In this direction,~\cite{zhou2017balanced} focuses on balancing the quantization values based on the distribution of the data. The quantizer can also be trained alongside the model~\cite{zhang2018lq,jung2019learning} and it is also possible to use reinforcement learning~\cite{wang2019haq} and meta learning~\cite{wang2020automatic} approaches to determine good choices for the quantizer.

    \textbf{Choosing quantized values.}
    There are various methods for quantizing data, ranging from simple heuristics like those used to convert network weights into binary values depending on their sign~\cite{courbariaux2015binaryconnect} or projecting real-valued parameters to (one of) the closest discrete points~\cite{jacob2018quantization}, to loss functions that regularize the network and force parameters into quantized states upon the convergence of the training algorithm~\cite{choi2020learning}.
    
    One notable approach is~\cite{zhou2017incremental}, which incrementally quantizes network weights to power of two terms. The set of non-quantized weights is progressively shrunk during re-training, with their values being updated to counter any accuracy loss induced by quantization. Knowledge distillation can also be a valid way to pick quantization values~\cite{bai2019few,polino2018model}.
    
    It is also possible to cast this task as a mathematical optimization problem. For instance,~\cite{lin2016fixed} converts pre-trained weights to fixed-point values by looking at their signal-to-noise ratio as an optimization metric. In~\cite{wu2020low}, the mean square error of the quantized data with respect to the original data is used to choose the precise 8-bit floating-point quantization format (mantissa and exponent size) and corresponding values. In more involved approaches, the Alternating Direction Method of Multipliers (ADMM) can be used to optimize the quantized values with low precision formats~\cite{chen2019deep,leng2017extremely}. Regularization terms and parameters that emphasize quantized solutions are also available. The work of~\cite{choi2020learning} looks at using mean squared quantization error regularization to drive weights to quantized values and how $\ell_2$ regularization can lead to sparse weight designs. Regularization also is an effective approach for doing robust quantization~\cite{alizadeh2020gradient,shkolnik2020robust}.

\subsection{Weight Sharing}
\label{subsec:weightsharing}

Weight sharing compresses the network by assigning shared values to parameters. This transforms plain weight data storage into a reduced number of shared values in a dedicated memory, together with the indices of these values in the weight matrix.


\begin{figure}[ht]
\centering
\includegraphics[width=0.75\linewidth]{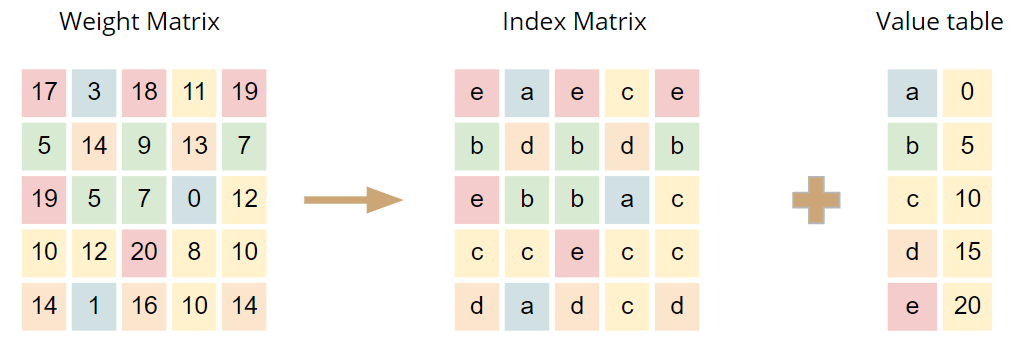}
\caption{Weight sharing techniques allow network compression by storing indices instead of values.}
\label{fig:ws_explanation}
\end{figure}

Figure~\ref{fig:ws_explanation} shows an example. The first matrix corresponds to a $5\times5$ convolutional kernel (filter) with values computed during training. The matrix contains $N=25$ values ranging from $0$ to $20$. Each value can be represented using $B=5$ bits, resulting in a total size of $N \cdot B = 25 \cdot 5 =  125$ bits. There are $5$ shared values, namely `a', `b', `c', `d' and `e', replacing the $25$ original values, as shown in the second matrix.


Accordingly, the size of an element in the weight matrix can be reduced from $B$ to $\log_2(K)$ bits, with $K$ being the number of different shared values. The size of the stored data then becomes $N\cdot\log_2(K) + K\cdot B$, instead of $N\cdot B$.

\begin{figure}[ht]
\centering
\includegraphics[width=0.75\linewidth]{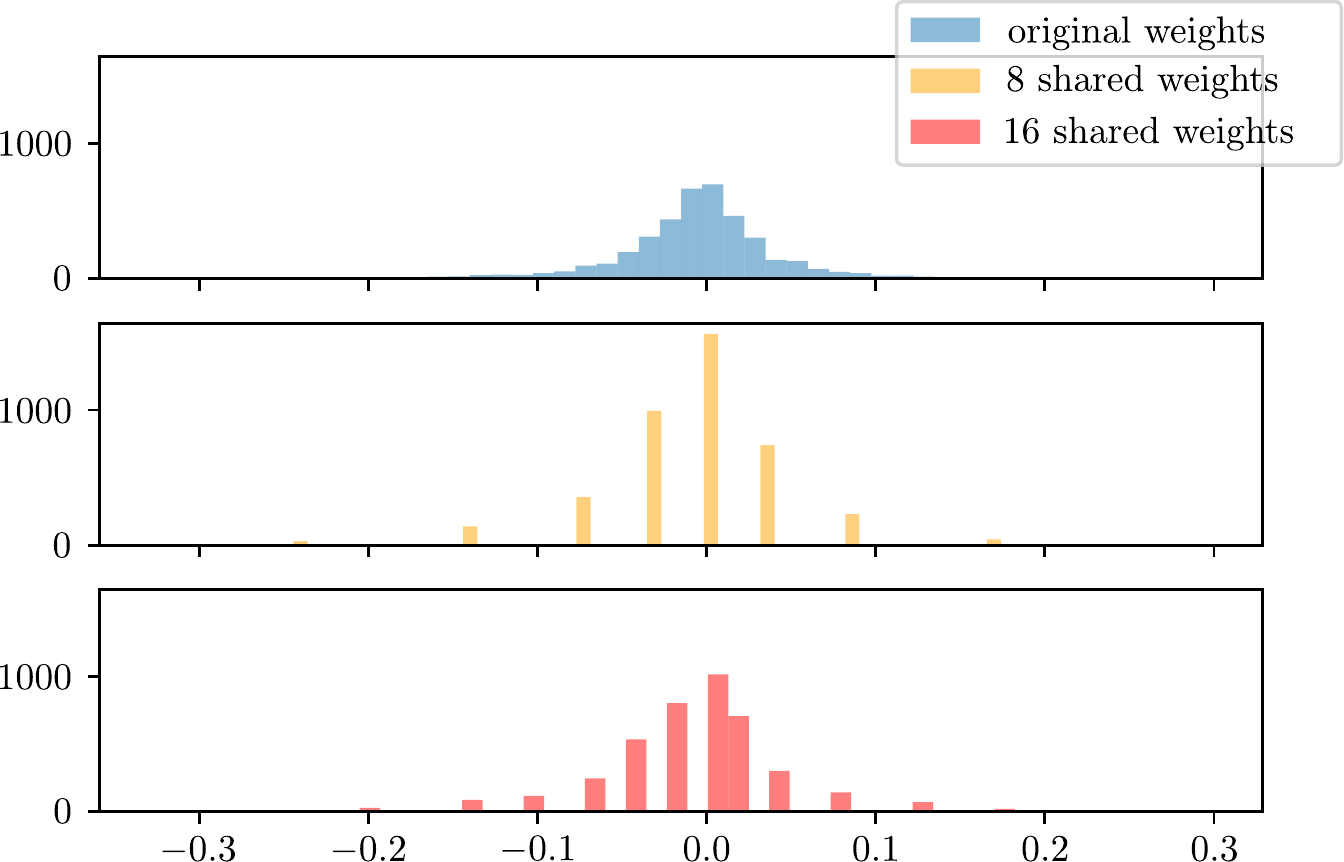}
\caption{Distribution of the weights composing the first layer of a trained ResNet50V2~\cite{ResNet}, original (top), with only 8 (middle) and 16 (bottom) shared values.}
\label{fig:ws_distribution}
\end{figure}

Depending on the number of shared values used, the distribution of the weights inside a layer will change. An example of this before and after weight sharing (with 8 and 16 shared values) can be found in Fig.~\ref{fig:ws_distribution}.


Weight sharing approaches can be classified by the method used to group weights together and by the granularity level it is applied at. Each of these aspects will be explained in some detail in the following paragraphs.

\textbf{Grouping methods.} One of the first approaches involving weight sharing that showed it can be a viable option for compressing neural networks is HashedNets~\cite{ChenWTWC2015hashednets}. The weights of the network in this setting are randomly grouped into hash buckets sharing the same value. These shared values are then trained and updated using backpropagation. The authors test their approach on the MNIST dataset with two custom fully connected networks with 3 and 5 layers.



However, instead of applying random grouping before the network even sees any data, it is also possible to approximate an already-trained network by determining groups based on weight values. In this vein, DeepCompression~\cite{han2015deepcompress} uses the K-means algorithm to iteratively group the weights in a network in a global 3-step compression approach involving network pruning, weight sharing, and parameter encoding. The K-means algorithm is used to cluster similar values together, followed by an iterative retraining phase. Different initialization options for the shared values are considered, with experiments showing that uniform initialization over the entire range of weight values works best. Applied to the AlexNet and VGG architectures on the ImageNet dataset, the compression algorithm achieves $35\times$ and $49\times$ compression, respectively, with negligible accuracy loss.


The most common way of doing K-means clustering is through the Lloyd algorithm~\cite{Lloyd82leastsquares}, which uses mean square error minimization to solve the clustering. However, this clustering approach does not imply performance loss minimization when taking into account quantization as well. The use of mean square error minimization does not necessarily lead to high accuracy during inference, even when uniform initialization of the clusters is used, as suggested with DeepCompression~\cite{han2015deepcompress}. Because of this,~\cite{choi2017ecqs-cnn} proposes to use Hessian-weighted K-means clustering to minimize accuracy loss. The approach consists of replacing the mean square error with the distortion of the Hessian matrix (second-order derivative) of the loss function. With this change, it can achieve a higher compression rate than DeepCompression, but with similar accuracy loss.


It is also possible to consider weight distribution when performing clustering. For instance,~\cite{Park2017WeightedEntropyBasedQF} proposes a clustering method based on weight entropy, using importance (magnitude) and frequency of the weights to group them. Thus, frequent non-zero (low importance) values are grouped, as well as rarer, but higher magnitude (high importance) values.

During the iterative process of training weights, clustering them, and training them again, previously clustered weights will sometimes diverge from the shared values at retraining time, making convergence to a good network model difficult. This is why, rather than applying iterative clustering and retraining,~\cite{wu2018deepkmeans} proposes the Deep-K-means approach that adds a regularization term in the training objective function, enforcing weights to stay clustered during training. After training is finished, the K-means algorithm is used to group the obtained weight values.


Other clustering algorithms can also be used. One main issue with using the K-means algorithm in this context is that it targets multi-dimensional data, whereas weights clustering is a 1-D problem. One example of approach using another clustering algorithm is DP-Net~\cite{yang2020dpnet}, which is based on a dynamic programming clustering algorithm that enables weight sharing in constant time, reducing the clustering complexity compared to the K-means algorithm.

\begin{figure}[ht]
\centering
\includegraphics[width=0.9\textwidth]{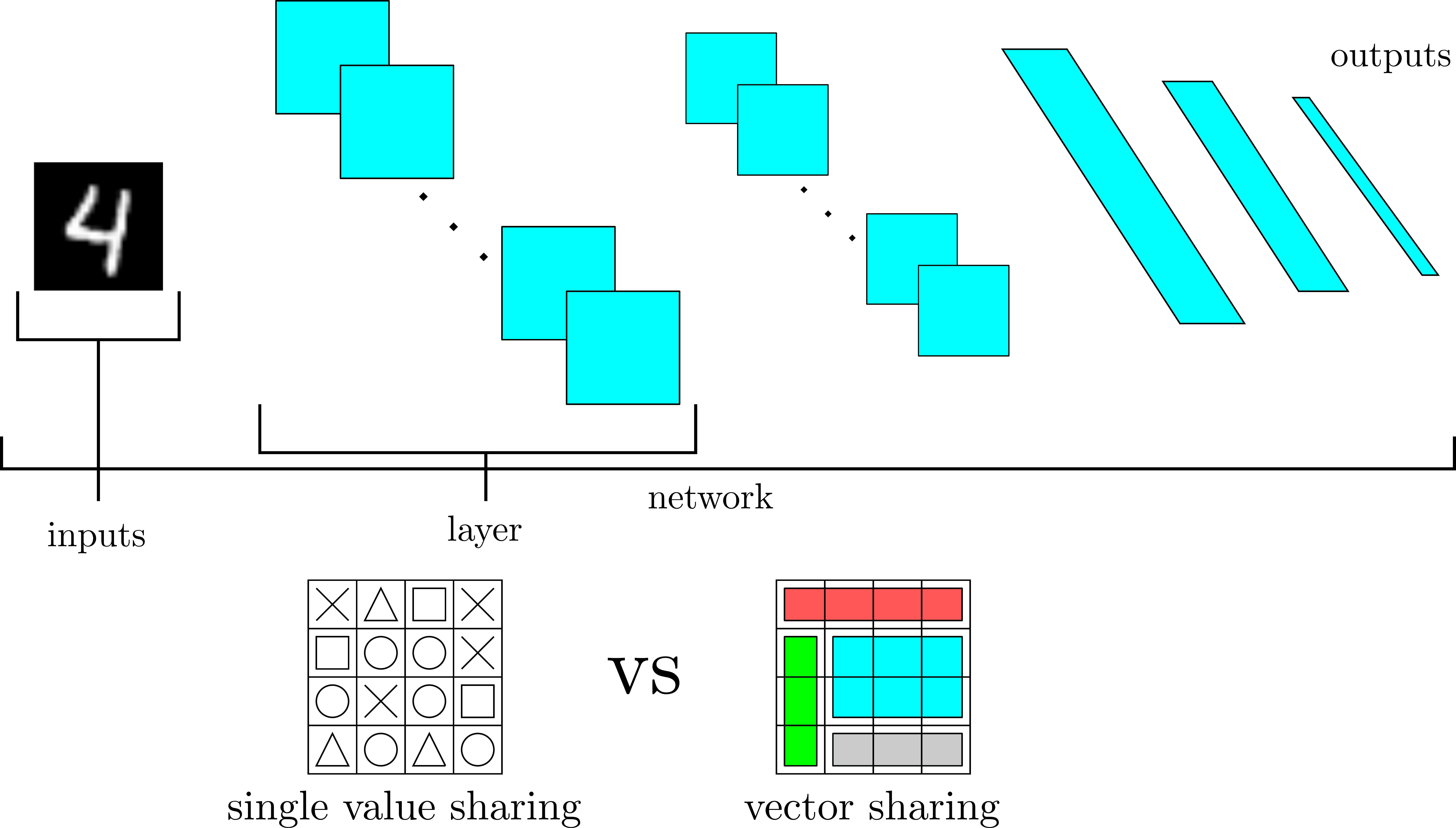}
\caption{The various scopes of applying weight sharing.}
\label{fig:scopeWS}       
\end{figure}

\textbf{Weight-sharing granularity.} The weights of a network can be shared at different levels of granularity, as shown in Figure~\ref{fig:scopeWS}. While this can be done for the entire network, as initially proposed in HashedNets~\cite{ChenWTWC2015hashednets}, each layer has a different weight distribution, covering a different range. Hence, sharing values for the whole network usually does not offer good enough representation power to limit accuracy loss.

On the other hand, sharing the values at the layer level offers a better representation of the original network, as shown with Deep Compression~\cite{han2015deepcompress}. Such an approach also allows different levels of compression to be used for each layer. The first and last layers are generally more sensitive to compression and require a higher number of shared values to keep accuracy loss acceptable. It is even possible to target a smaller scope, like sharing values at the level of a (convolutional) kernel -- but then of course the compression rate will be much lower.

While reducing the scope allows a better representation of the initial weight distribution, thus keeping accuracy loss low, it is possible to improve compression performance. For example, Deep-K-means~\cite{wu2018deepkmeans} shares values at a level that is optimal for the very efficient row-stationary dataflow used in DNN hardware accelerators.

Even if weight sharing leads to good compression rates, it does not enable inference acceleration by itself. This can be achieved if inputs are also discretized, reducing the number of combination operations and allowing the use of a pre-computed look-up table multiplier. This approach is used in LookNN~\cite{razlighi2017looknn}, which applies K-means to the input feature map to achieve a nonlinear quantization whereas the remaining feature maps are quantized in the traditional linear way.

Values can also be shared at a smaller level, as in Q-CNN~\cite{wu2016q-cnn}. Here, layers are decomposed into sub-vectors, which are then clustered using the K-means algorithm. 
Sharing vectors like this reduces the number of possibilities when performing products. This enables the layer response to be approximated using product pre-computation with a look-up table.


\subsection{Network Sparsification (Pruning)}
\label{subsec:pruning}


DNNs tend to be more complex as their accuracy rate improves and this complexity usually carries with it the fact that the network is over-parameterized. On the other hand, it has been argued for a long time~\cite{cun1990optimalbrain} that structure is more important than density in neural networks, with sparse models having the ability to generalize up to as well as their dense counterparts. Removing model parameters has the direct effect of reducing the size of the model, but it can also be used for speeding up the inference process by reducing the number of computations. Depending on the objective, different parts of the network can be more interesting to prune than others. For instance, fully connected layers usually concentrate most of the network weights in a CNN and should be targeted for high compression. Convolutional layers, however, contain fewer model parameters but account for most of the computations. Since they generate the majority of data movement in the model, they should be targeted when model performance and energy efficiency are important.


Pruning methods can be classified by how they are applied to the network, the granularity of the pruning, and finally the saliency determination approach. All these criteria are discussed in the following paragraphs.

\textbf{Target regions.} 
The loss in accuracy incurred by removing parameters can be recovered by re-training the remaining parameters using the initial training dataset if it is still available. This pruning process can be performed at different steps of the network life-cycle, either prior, during, or after training the model.

    It has been shown that some parts of DNNs are more resilient to approximation than others. As such, pruning each layer at the same rate is not very efficient for accuracy. But at the same time, choosing the optimal sparsity level for the whole network is a complicated task. For example,~\cite{he2018amc} proposes to heuristically optimize the pruning ratio of each layer using reinforcement learning.

    Similar to pruning weights, feature maps can also be pruned during the forward pass of the network. This process is called~\emph{dynamic sparsity} and is used in many accelerators to avoid zero or near zero computations~\cite{huan2016nearzero,huan2017enlargednearzero}. Such approaches require dedicated architectures, but since the focus is only on data type refinement methods for this survey, they will not be discussed further.
        
\textbf{Pruning Granularity.}
Depending on the pruning objective (compression or performance), one can choose to focus on weight removal at various sparsity levels. For instance, even though removing an entire structure (\emph{e.g.,}~a convolution kernel) allows reducing the computational complexity of the model, and thus, improving performance, it also has the effect of inducing a higher accuracy loss.
\
        
\begin{figure}[ht]
\centering
\includegraphics[width=0.8\textwidth]{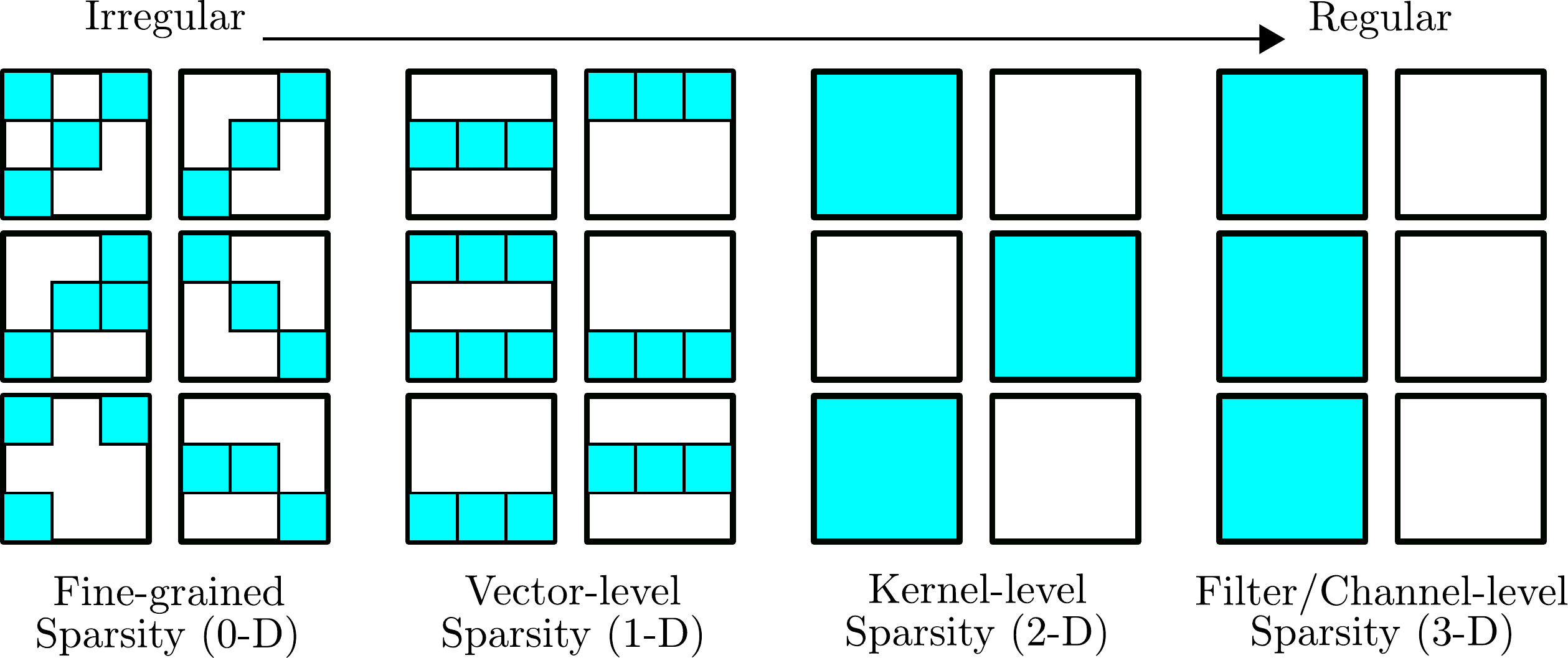}
\caption{Different granularities of pruning in a 4-dimensional weight tensor for DNN inference (adapted from~\cite[Figure 1]{mao2017exploring}).}
\label{fig:granularityPruning}       
\end{figure}
        
The lowest pruning level is at the weight level, the goal being that of removing the individual parameters with the lowest saliency~\cite{cun1990optimalbrain,han2015deepcompress}. Although this generally results in the lowest accuracy loss, it does not systematically offer latency or energy improvements because sparse tensor computations are quite difficult to accelerate. Its main purpose is therefore to compress the network in memory. 
        
To accelerate computations, a regular sparsity pattern is usually required. This is called~\emph{structured pruning} and aims at removing (spatially close) groups of weights so that network inference can be simplified. To achieve this,~\cite{Ji2018TETRISTT} iteratively reorders pruned weights to prune larger structures, whereas~\cite{Yu2017ScalpelCD} uses different pruning strategies depending on the hardware, optimizing for the full utilization of available SIMD units.


As previously hinted, it is also possible to remove convolution kernels, thus simplifying the processing of pruned convolutional kernels. An example is~\cite{Molchanov2016ressourceefficientpruning}, which progressively removes convolutional kernels through greedy-based fine-tuning. The method is applied to transfer learning applications, resulting in a $2\times$ speedup on ImageNet-class CNNs.
        
Another interesting structure amenable for removal is a channel. Once channels are removed, one can remove the corresponding filters that take these channels as input. The filters producing these channels in the previous layer can also be removed~\cite{Luo2017ThiNetAF}. A representative approach is~\cite{he2017channelpruning} which removes channels based on importance, resulting in a $2-5\times$ speedup on multiple ImageNet-class CNNs with under $1\%$ accuracy loss. In subsequent work,~\cite{he2018amc} proposes to pick the pruning ratio of each layer using reinforcement learning.
        
\textbf{Weight Saliency Determination.}
Removing part(s) of the network usually requires knowing which regions are least important for ensuring network accuracy. This is called \emph{saliency determination} and it can be conducted using different methods, as described next. A simple way is to use heuristics like weight magnitude or examining the $\ell_1/\ell_2$ norm of a group of weights, whereas more recent work employs optimization algorithms to address the trade-offs between accuracy loss and compression/acceleration.


The earliest methods removed small magnitude weights because they tend to have the least impact on accuracy~\cite{cun1990optimalbrain,Hassibi1992SecondOD}. They work iteratively by fine-tuning unpruned weights to recover lost accuracy~\cite{han2015deepcompress}. It has been shown recently that one can also remove redundant connections in FC layers since for weights having the same value, only one needs to be kept~\cite{Srinivas2015DatafreePP}. If accuracy is degraded too much during the pruning process, some methods can be used post-pruning to restore certain weights and improve accuracy~\cite{Guo2016DynamicNS,Narang2017ExploringSI}. For convolutional layer filter removal, it is possible to rank filters based on their $\ell_1$ norm and prune the lowest ranking filters of each layer~\cite{Li2017PruningFF}. Instead of ranking filters at the layer level, one can also do it at a global, network-wide level by first doing a layer-wise filter ordering using $\ell_2$ norms and then computing affine mappings that enable inter-layer filter rankings~\cite{Chin2020learnedglobalranking,Ting2016LeGR}. Such global approaches lead to a Pareto set of approximated networks that offer various trade-offs between performance and accuracy.

In~\cite{Molchanov2016ressourceefficientpruning}, the authors also consider a Taylor expansion criterion that approximates accuracy degradation due to feature map removal. This is done using activation and gradient values already computed during a regular training iteration. Other approaches use weight gradients to compute saliency. For instance,~\cite{Dai2017NeST} proposes a sequential two-step process where (1) gradient-based information is used to grow the network (adding `dormant' connections and neurons that are deemed important for accuracy) and (2) regular magnitude-based pruning of weights and connections.
        
Another method to identify representative structures inside a network is~\cite{he2017channelpruning}, which uses a two-step process involving Least Absolute Shrinkage and Selection Operator (LASSO) regression for channel selection and then a least squares-based reconstruction approach of subsequent feature maps in the network.
    


It is also possible to state the problem of selecting which parts of the network to remove as an optimization problem. One example is~\cite{Luo2017ThiNetAF}, which relies on the correlation between feature maps of the current layer and the next one to determine the importance of filters. In another approach~\cite{Yang2017DesigningEC}, the optimization problem features the model's energy efficiency as an objective. It is based on an energy estimation methodology capable of approximating both the power of MAC operations and data access (which is more complicated to compute, depending on the data reuse technique). The resulting iterative process involves local fine-tuning to recover accuracy loss in a layer before moving on to subsequent layers.


By formulating weight pruning as a non-convex optimization problem, it is possible to address it using an ADMM approach~\cite{Zhang2018ASD}. Using the desired sparsity level as a constraint to be satisfied and the loss of the network as the objective to minimize, ADMM can be used in a two-step process. Since convergence can be quite slow, the target error is increased to accelerate convergence and the resulting accuracy loss is compensated by network retraining. The method can also be extended to address high sparsity target problems, by introducing a more progressive algorithm using partial weight pruning with a moderate pruning rate~\cite{Ye2018ProgressiveADMM}.
        
Another idea is to encourage weights to group around zero using regularization. The closer weights are to zero, the less accuracy loss will be induced by removing them. For example,~\cite{Lebedev2016FastCU,Wen2016LearningSS} used group LASSO~\cite{Yuan06modelselection} regularization to obtain structured sparsity, with the same factor being applied to all the weight groups. In~\cite{Liu2017NetSlimming}, $\ell_1$ regularization is applied to the scaling factor of batch normalization layers to identify important channels. Different regularization factors can be assigned to different groups, such as in~\cite{Ding2018AutoBalancedFP}, where $\ell_2$ regularization is used to transfer the model's representational capacity to a fraction of its filters. An incremental approach for choosing these factors can also be used~\cite{Wang2019StructuredPF}. In~\cite{Luo2020AutoPrunerAE}, feature map channels are gradually zeroed during training using a dynamic regularization factor (whose value depends on the current compression ratio in the network), allowing safe removal of corresponding filters without a significant drop in the accuracy.
        
Another recent approach to optimize pruning is through architecture search. Usually, pruning methods target a fully trained network and recover any accuracy loss using fine-tuning because it is hard to train a sparse network. Recently, however, the idea that a classic network contains sub-networks that, trained from scratch, can perform as well as the original network but with fewer parameters and computation, was introduced~\cite{Frankle2018LotteryTickets}. This idea was also explored in~\cite{Liu2019RethinkingTV}, which claims that directly training (using some form of random initialization) a model found at the end of a classic three-step pruning process (training, pruning, and fine-tuning) can perform as well, if not better, in fewer training steps. The issue is that, in the beginning, none of these studies provided a method for finding an efficient smaller architecture without doing full model training beforehand. This is starting to change, with~\cite{Lin2020ChannelPV} proposing to use a bee colony exploration algorithm to find an appropriate DNN pruning scheme. It is also possible to reduce the fine-tuning cost by using an external network trained to predict weights of a certain network structure, facilitating a fast exploration of various possible architectures~\cite{Liu2019MetaPruningML}.
\section{Approximation for Training}
\label{sec:training}

The state-of-the-art models used in deep learning applications require a considerable hardware infrastructure to be designed properly. There are various challenges related to computing, storage, network/communication, as well as memory capacity and bandwidth that can potentially hinder the scalability of current solutions to future models and applications. This is most visible during the training part of neural network design, which accounts for the majority of the computing time and resources. 



Accelerating training at the arithmetic level has thus become a hot research topic,
but early work in this direction did not necessarily translate to a wide adoption 
and availability of low/mixed-precision training hardware. For example, BinaryConnect~\cite{courbariaux2015binaryconnect} introduced a CNN training 
methodology with binary ($+1$ and $-1$) weights, with all other operations 
and data structures (\emph{e.g.,}~tensors) in full \texttt{float32} precision. 
This binarization was soon extended to include activations~\cite{courbariaux2016binarizedb}, 
followed by experiments with quantization levels of $2, 4$ and $6$ bits for weights and activations~\cite{hubara2017quantized}, but with backpropagation gradients still computed and stored in full precision. Binarization for all tensor operations, including gradient computations, is considered in XNOR-Net~\cite{rastegari2016xnor}. While ensuring impressive efficiency gains, these approaches lead to non-trivial accuracy loss for larger CNN models that have since been introduced and adopted in practice. 

To manage accuracy loss, DoReFaNet~\cite{zhou2016dorefa} uses different quantization
bit widths for weights, activations, and gradients, but still incurs some
accuracy loss and requires exploring different bit width configurations on 
a per-network basis, which can be impractical for large models. The approach 
introduced in~\cite{mishra2017wrpn} improves on previous accuracy results 
by doubling or tripling the number of inputs and outputs of layers in popular CNN models, but 
again requires that gradients be computed and stored in full precision and 
does not achieve the same accuracy as the baseline non-quantized trained model.

Studies with fixed-point arithmetic on DNNs have also been conducted since the early 
1990s~\cite{holt1991back,presley1994fixed,simard1994backpropagation, 
savich2007impact} and more recently~\cite{gupta2015deep} has shown that a 16-bit 
fixed-point representation coupled with stochastic rounding can be used to 
train CNNs on the MNIST and CIFAR-10 datasets without accuracy loss. Nevertheless,
it is unlikely that this approach would work on larger CNNs trained on 
larger datasets.

There have also been several proposals for quantizing recurrent neural network (RNN) training. For 
instance, in~\cite{he2016effective}, training for quantized versions of 
gated recurrent units and long short-term memory cells with few bits for weights and activations are investigated,
with a slight loss in accuracy with respect to base full precision models.
A different approach~\cite{ott2016recurrent} evaluates binary, ternary and 
exponential quantization for weights used in various RNN models trained 
for speech recognition and language modeling. Similar to the CNN-centered
methods evoked so far however, all these approaches use full precision gradients, 
and therefore do not improve computation cost during backpropagation.


\subsection{Mixed Precision Training approaches}\label{sec:mpt}
The most widespread approach to increase performance and efficiency of 
DNN training at the arithmetic level is through the use of~\emph{mixed 
precision}. 

On the commercial side, NVIDIA has offered the possibility 
to do low precision training since the Pascal architecture in 2016 and 
mixed precision training (combining \texttt{float16} and \texttt{float32} 
arithmetic) has really taken off with the subsequent introduction of 
TensorCore units in their Volta and Turing architectures in 2017--2018. 
TensorCores are, in essence, programmable $4\times 4\times 4$ 
matrix-multiply-and-accumulate units (performing the operation
$D = A\times B + C$, where $A, B, C$ and $D$ are $4\times 4$ matrices,
with $A$ and $B$ stored using \texttt{float16} and $C$ and $D$ being
either \texttt{float16} or \texttt{float32} matrices). An execution of
a large number of such units provides a huge performance 
boost (several times when compared to NVIDIA's previous Pascal hardware) 
to convolution and matrix operations with mixed precision operands 
and results. Over at Google, their newer (from version V2
onward) Tensor Processing Units (TPUs) offer similar support for 
mixed precision training with the introduction of \texttt{bfloat16}, 
a 16-bit floating-point format that, when compared to \texttt{float16}, 
trades in mantissa bits for exponent bits (a $5$-bit exponent and 
$10$-bit mantissa for \texttt{float16} versus an $8$-bit exponent and 
$7$-bit mantissa for \texttt{bfloat16}).
Intel and ARM are also adopting \texttt{bfloat16} in their push to 
offer AI-enhanced hardware, while AMD has introduced software support 
for \texttt{bfloat16} in recent versions of 
their ROCm platform.
As of May 2020, the Ampere architecture from NVIDIA also introduces
\texttt{bfloat16} operator support in their third version of TensorCore
units.


\subsubsection{Mixed 16-32-bit precision training}\label{sec:16_32_mpt}
An important remark about backpropagation training that should guide 
the choice of number formats is how the values contained within
various quantities (activations, gradients, and parameters) vary during
successive training iterations. It is noted in~\cite{courbariaux2014training}
that ``activations, gradients and parameters have very different ranges'', whereas
``gradient ranges slowly diminish during training''. 
There is also the idea that a higher numerical precision should be used when updating the parameters than when using them during the back and forward propagation
operations~\cite[Sec.~6]{courbariaux2014training}. Recent accelerated training approaches (at the arithmetic level) follow these observations. 

\begin{figure}[ht]
    \centering
    \includegraphics[width=0.9\linewidth]{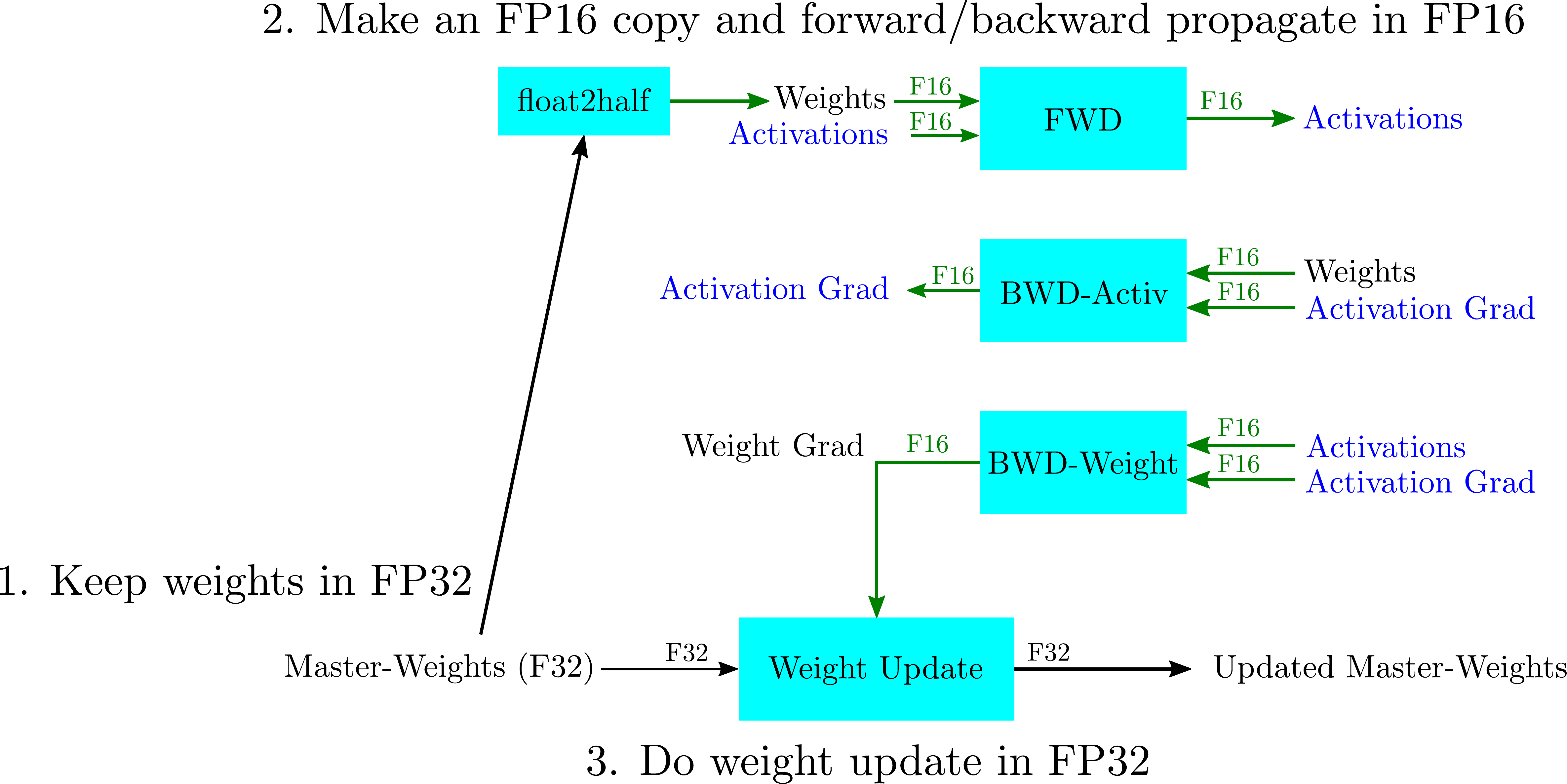}
    \caption{Mixed precision training iteration for a network layer 
    (adapted from~\cite[Fig.~1]{micikevicius2017mixed}).}
    \label{fig:mpt_iteration}
\end{figure}

\textbf{An approach for} \texttt{float16}\textbf{-based training acceleration.}
In~\cite{micikevicius2017mixed}, NVIDIA TensorCores are used to perform 
mixed \texttt{float16} and \texttt{float32} operations during each training
iteration. The process is illustrated in Figure~\ref{fig:mpt_iteration}: a 
full precision copy of the weights is always stored and updated at each 
iteration, whereas the gradient computations of the weights and activations 
are done using \texttt{float16} quantizations of the weights. The dot-product 
and reduction (\emph{i.e.,}~sums of elements across a vector) operations 
are performed with a~\texttt{float32} accumulator (as is enabled by
TensorCores), which, according to~\cite{micikevicius2017mixed}, is needed 
in some cases to maintain the same model accuracy as with a 
baseline~\texttt{float32} approach.

The main reason for using 32-bit values for the weight updates is that during 
later iterations of training, the update gradients become too small to be 
used with \texttt{float16} addition, which will result in them getting 
clipped when $\bs{w}^t \gg \varepsilon\frac{\partial \ell}{
\partial \bs{w}^t}$ and adversely affect the final model accuracy. For 
\texttt{float16}, this happens when the ratio between weight and update 
is larger than $2048$.

\begin{figure}[ht]
    \centering
    \includegraphics[width=0.7\linewidth]{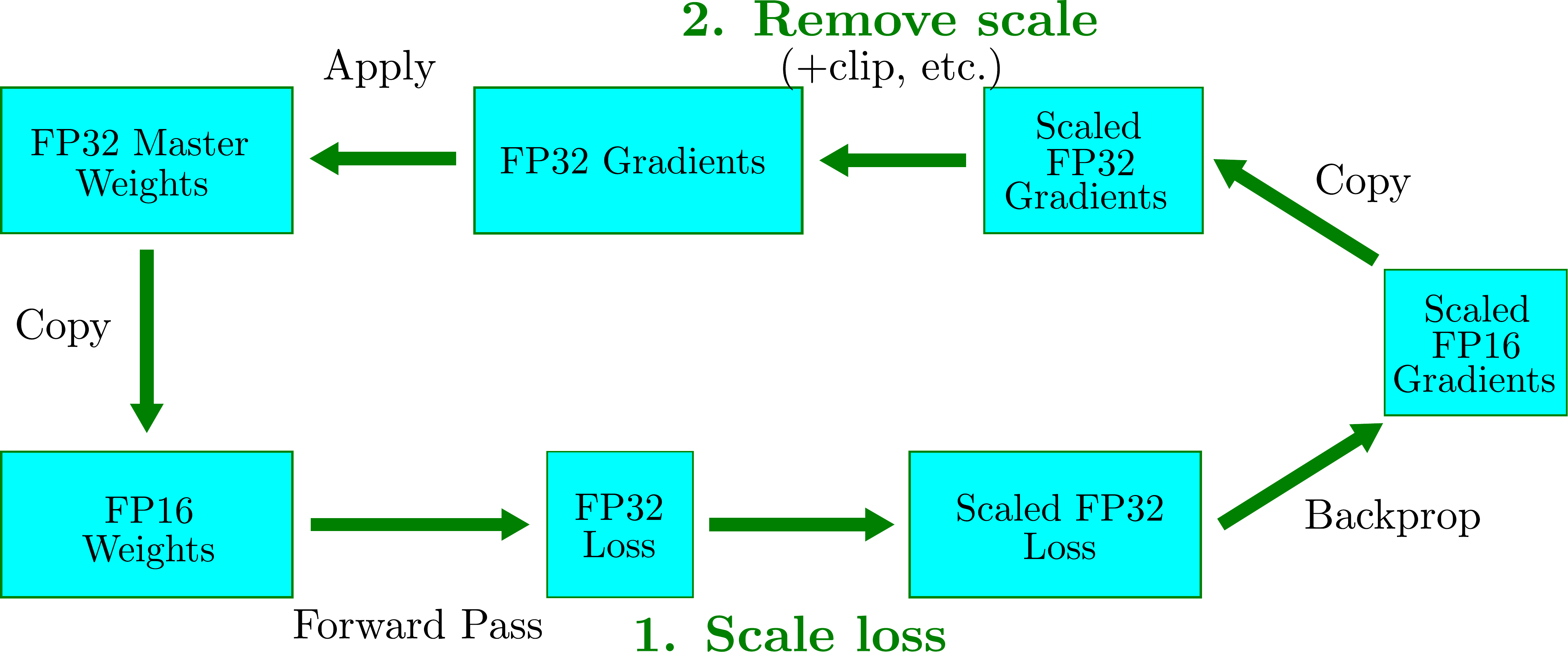}
    \caption{The loss scaling procedure for updating the master weights in 
    mixed precision training.}
    \label{fig:loss_scaling}
\end{figure}

A related issue when gradients become too small is that they might not be 
accurately representable in \texttt{float16}, even though the dynamic range 
of the weight/activation gradients at each layer is much smaller than the 
$2^{40}$ range associated with \texttt{float16}. This means that a scaling
approach might be applicable. This is indeed what is advocated in~\cite{micikevicius2017mixed},
where gradient values can be shifted to \texttt{float16}-representable ranges by scaling the loss value computed during the forward pass, before performing backpropagation. By chain rule calculus during backpropagation, all gradient
values will then be scaled by the same amount. Weight gradients will have to be unscaled back before weight update to ensure the same update process
as with \texttt{float32} training. The entire procedure is summarized in 
Figure~\ref{fig:loss_scaling}. Although not explored in~\cite{micikevicius2017mixed}, the
scaling factor can be chosen automatically: 
start with a very large scaling factor (\emph{e.g.,}~$2^{24}$), if gradient 
overflows (with \texttt{Inf} or \texttt{NaN}) decrease the scale by a factor of $2$
and skip the current update, whereas if no overflow has occurred for some time 
(\emph{e.g.,}~$2000$ iterations), increase the scale by a factor of $2$.

The results presented in~\cite[Sec.~4]{micikevicius2017mixed} show that mixed-precision 
training is a viable alternative (in the sense that it gives comparable results 
to baseline \texttt{float32} training) for various tasks such as image classification
(with tests on AlexNet, VGG-D, GoogLeNet (Inception v1), Inception v2 \& v3 and ResNet50),
object detection, speech recognition, machine translation, language modeling and 
Generative Adversarial Networks (GAN) generation.

In addition to the speed benefit that such a mixed-precision training approach 
brings (which varies from $2\times$ to $6\times$ with respect to baseline 
training on the experiments carried out in~\cite{micikevicius2017mixed} on a
Volta GPU), the memory consumption for training is roughly halved, since the 
dominating quantities are the activations (due to larger batch sizes and the 
fact that they need to be stored for reuse during back-propagation), which 
are stored in~\texttt{float16}.

\textbf{Enabling} \texttt{bfloat16}\textbf{-based training methods.}
It seems that the need for loss scaling can be avoided if the~\texttt{float16}
format and associated operations are replaced with 
\texttt{bfloat16} (this is shown in~\cite{kalamkar2019study}, where experiments 
with various state of the art networks in image classification, speech recognition,
language modelling, generative networks and industrial recommendation systems
show the versatility of \texttt{bfloat16}-based training). 
This is due to the fact that~\texttt{bfloat16} has the same exponent range 
as~\texttt{float32} and the lower mantissa width does not adversely impact the final model accuracy. There are also additional hardware-related benefits 
that come with the combination of~\texttt{bfloat16} and~\texttt{float32}. Core computational primitives such as FMA units can be built using 8-bit multipliers, 
leading to a significant area and power savings while preserving the full dynamic range of~\texttt{float32}.

The appeal of using~\texttt{bfloat16} is that it also does not require any 
changes to the training model (as designed for a baseline \texttt{float32}
approach). The increasing (planned) hardware support from several vendors seems to suggest it will soon be the~\emph{de facto} choice for performing DNN training, replacing the aforementioned~\texttt{float16} approach.
This statement is strengthened by the added support of~\texttt{bfloat16} on 
NVIDIA's Ampere GPU architecture.

\textbf{Fixed-point-based training.}
Mixed precision training approaches that are based mostly on integer/fixed-point 
arithmetic has also been proposed recently. These methods~\cite{courbariaux2014training,
koster2017flexpoint,das2018mixed,drumond2018training} use during computation integer tensors with tensor-wide shared exponents. The format explored in~\cite{courbariaux2014training}
has an 11-bit mantissa and a 5-bit shared exponent, tested on custom maxout~\cite{goodfellow2013maxout} 
networks for the MNIST, CIFAR-10, and SVHN datasets. At each layer, every weight, bias,
activation input \& output, gradient vectors, and matrices have different exponent values.
These exponents are updated based on a~\emph{passive} over/underflow detection policy
which is run periodically during training. Because it is just reacting to the presence of overflows in the networks, it can potentially impede convergence of the training process.

To address this problem,~\cite{koster2017flexpoint} proposes widening the dynamic
fixed-point format to a 16-bit mantissa and a 5-bit shared exponent, a format which 
they call \texttt{flexpoint} (\texttt{flex16+5}). They also introduce a new 
algorithm (Autoflex) for adjusting the shared exponents in an~\emph{adaptive}
the way each time a tensor is written to, using tensor-wide statistics gathered 
at previous iterations. This essentially eliminates the appearance of overflow 
errors, leading to results on par with baseline \texttt{float32} training 
on AlexNet, ResNet-110 and Wasserstein GAN models. Choosing the bit widths 
that resulted in the~\texttt{flex16+5} format was done such that the mantissa 
can encode most of the variability of values inside a tensor during one training 
epoch and that for weight update operations there will be sufficient
mantissa overlap between tensors to ensure accurate computation (which seems to
eliminate the need for 32-bit master copies of the weights during the update
process).

The Flexpoint approach would require the presence of dedicated hardware for it to truly show its effectiveness. That is why in~\cite{das2018mixed} another dynamic fixed-point representation that can leverage already existing
general-purpose hardware (through the use of existing integer operations)
is presented. The mantissa is again 16-bit, while the shared exponent is stored as an 8-bit integer. The matrix multiply and dot product operations 
needed for the training procedure are done using 16-bit input 32-bit output 
integer FMAs, with some intermediate accumulations converted to \texttt{float32}
in order to avoid overflows in long addition chains. Similar 
to~\cite{micikevicius2017mixed}, a \texttt{float32} master copy of the weights 
is kept at each iteration for the update process. Tests are carried out on 
Intel XeonPhi Knights-Mill hardware for several CNN models (ResNet-50, 
GoogLeNet-v1, VGG-16 and AlexNet) on ImageNet, showing an $1.8\times$ 
speedup over baseline~\texttt{float32} training on the same platform.

While using tensors with shared exponents can lead to performance and 
efficiency gains in the just discussed methods,~\cite{drumond2018training} 
identifies three potential roadblocks in their use for training acceleration: 
(1) whereas dot product operations can be area-efficient with such formats, 
other operations might be less efficient; (2) exponent sharing can lead to 
data loss if magnitudes are too large or too small, making exponent selection 
critical; (3) data loss can happen if the tensor value distributions are too 
wide to be captured by the allotted number of mantissa bits. To address 
them,~\cite{drumond2018training} proposes a hybrid approach, where all dot 
product operations are performed with shared exponent formats, while other 
operations are kept in floating-point. Since training operations are dominated 
by dot products, there will be little overhead to using floating-point for 
the remaining operations.

By using tiling for matrix multiplications (with shared exponent at tile level)
and wider weight storage for the weight update process (similar to other 
approaches),~\cite{drumond2018training} can limit data loss when 
compared to baseline \texttt{float32} training on a large range of tasks,
with little silicon density penalty. Investigating the design space, they 
find that the hybrid approach is most convenient for $24\times 24$ tile sizes,
$8$ to $12$-bit mantissa and $16$-bit size for weight storage.

\subsubsection{Mixed 8-16-bit precision training}\label{sec:8_16_mpt}
While combined 16-32-bit training seems to be the most widespread approach currently, for accelerating DNN training, there has also been work recently to 
push the envelope further with 8-bit tensor datatypes and multiplication operators
coupled with 16-bit accumulators and weight updates~\cite{wang2018training,
mellempudi2019mixed} (instead of the 16-32-bit mix advocated in 
Section~\ref{sec:16_32_mpt}).

According to~\cite[Sec.~1]{wang2018training}, there are three main elements that
can significantly impact model test accuracy when using extremely low 
precision formats during training: (a) all operands in a tensor matrix multiply 
operations (GEMMs and convolutions) are in 8-bit formats (2\% degradation over 
a baseline~\texttt{float32} training loop on ResNet18 with the ImageNet dataset),
(b) GEMM accumulation results reduced from 32 to 16 bits (while critical to 
reducing the area and power of 8-bit hardware, such a move also leads to 
significant degradation --- 1\% with respect to the same ResNet18 baseline)
and (c) reducing weight updates from 32 to 16-bits (high precision weight 
updates and gradients require expensive parameter copies to be kept in memory,
whereas reducing their precision can also lead to significant degradation ---
1.7\% with respect to the ResNet18 baseline).

To cope with these problems,~\cite{wang2018training} advocates the choice of 
a 5-bit exponent and 2-bit mantissa floating-point format to represent 
weights, activations, errors, and gradients in matrix multiply operations (forward, 
backward and gradient), coupled with a 6-bit exponent 9-bit mantissa format for 
all the accumulation results. These format choices are motivated by 
how data is distributed inside networks in practice, with a focus on striking 
a balance between representation accuracy and dynamic range. To optimize the 
accuracy of the accumulation, a blocked approach (which is standard in high 
performance basic linear algebra routines) is used. The multiplications are done in the 8-bit format,
whereas the accumulation is done in 16-bits to more accurately model the result (\emph{i.e.,}~try
to avoid \emph{stagnation}/\emph{swamping} from appearing: small $x_ky_k$ terms cannot 
contribute to $\sum_{k=1}^{n}x_ky_k$ in the floating-point computation path).

\begin{figure}[ht]
    \centering
    \includegraphics[width=0.9\linewidth]{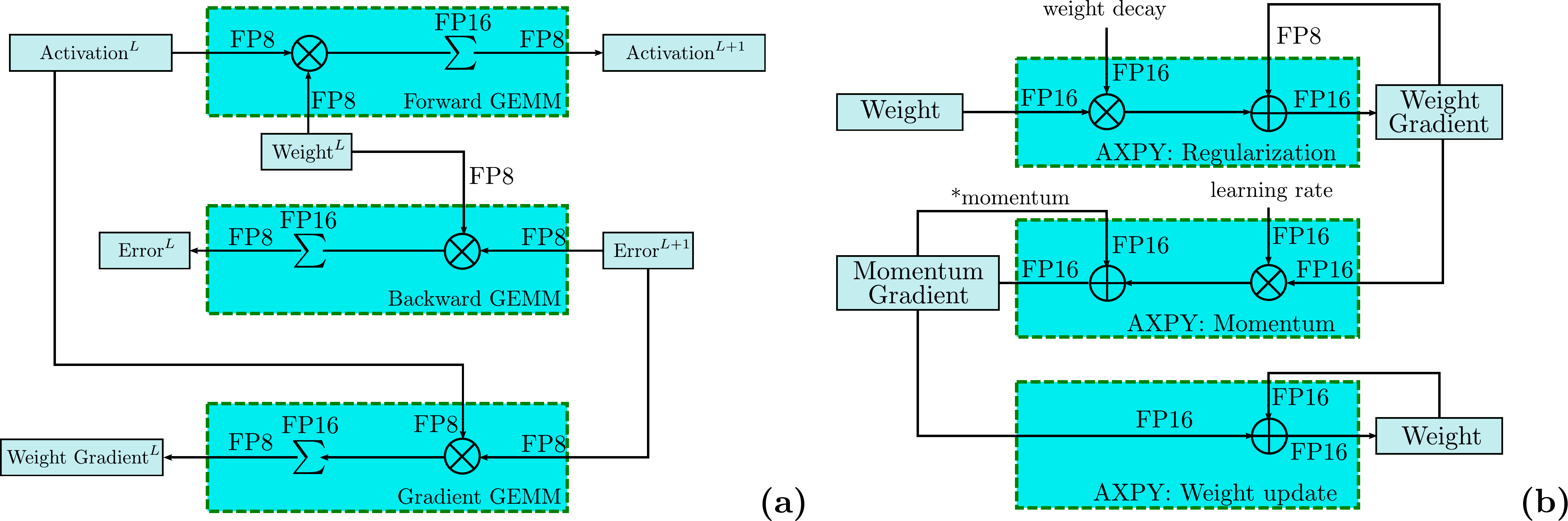}
    \caption{Summary of the precision settings for (a) the GEMM operations 
    during the forward and backward passes in backpropagation and (b)
    the AXPY operations during a standard SGD weight update process
    (adapted from~\cite[Fig.~2]{wang2018training}).}
    \label{fig:fp8training_1}
\end{figure}

Another way to improve on the overall accuracy of summation results is to use stochastic 
rounding, which shows similar results to block accumulation (see~\cite[Fig.~3]{wang2018training}).
In the context of deep learning, it seems that using stochastic rounding is more natural
for the weight update process (in the dot product AXPY operations) since the weight gradient 
is accumulated into the weight over mini-batches during several epochs (so not at once 
in a complete dot product operation!). 

The precision settings for all the operations done during training are summarized in Figure~\ref{fig:fp8training_1}. In terms of results, a large spectrum of neural 
networks for both image classification and object recognition are used (AlexNet and 
ResNet 18 and 50 versions for the ImageNet and CIFAR10 datasets) with both SGD 
and ADAM-based optimizers. A loss scaling approach similar to~\cite{micikevicius2017mixed}
is used to preserve the dynamic range of back-propagated errors with small magnitude.

In both~\cite{micikevicius2017mixed,wang2018training}, the hardware complexity of 
the floating-point computation pipeline is dominated by the accumulator bandwidth
(32 \& 16 bit, respectively), and in many cases, this size seems much too conservative. The follow-up work~\cite{sakr2019accumulation} introduces an 
analytical method for predicting the precision requirements for partial sum 
accumulation in the three GEMM accumulation units from Figure~\ref{fig:fp8training_1}.
It studies in what (precision/format) scenarios the variance of the accumulator units is maintained when doing dot product computations in reduced precision.

\begin{figure}[ht]
    \centering
    \includegraphics[width=0.7\linewidth]{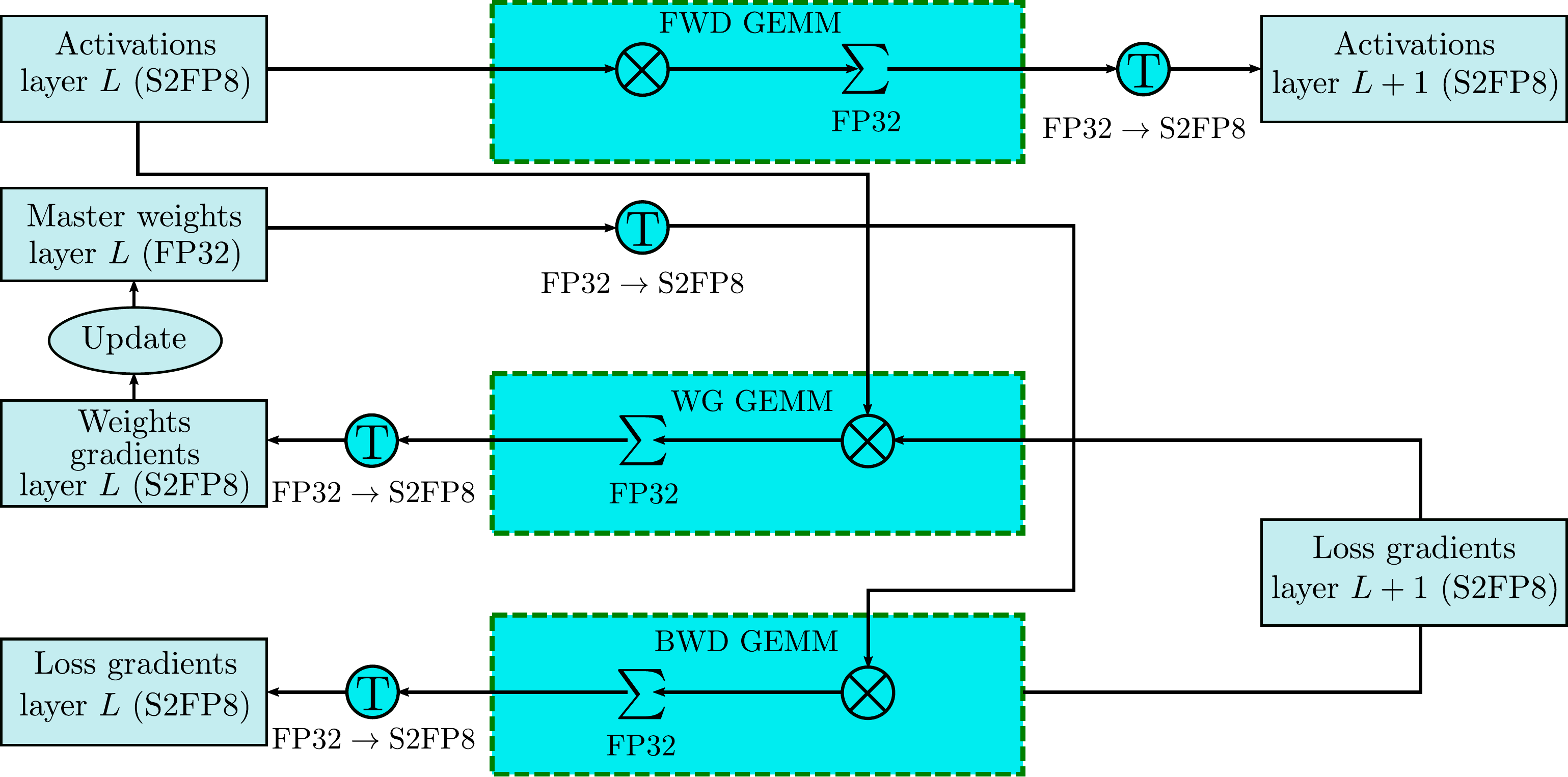}
    \caption{The low precision training flow with the S2FP8 format, where 
    the truncation function $T$ corresponds to $T(X)=\left[2^{-\beta}\left\lbrace
    \text{round}_{\text{FP8}}(2^\beta|X|^\alpha)
    \right\rbrace\right]^{1/\alpha}$. The forward and backward GEMM 
    operations use only S2FP8 values, whereas the weight update step uses
    FP32 master weights (adapted from~\cite[Fig.~4]{cambier2020shifted}).}
    \label{fig:fp8training_3}
\end{figure}

One downside of all these aforementioned methods is that they require certain 
knobs to be finely tuned (such as appropriate chunk-based accumulator design,
stochastic rounding techniques, loss scaling, and maintaining some layers of 
the network in higher precision -- in particular the first and last ones), 
necessitating experimentation on a network-by-network basis. To eliminate 
the need for such fine-tuning,~\cite{cambier2020shifted} proposes a new, 
tensor-level 8-bit floating-point format. Given an $N$-element tensor 
$X=\left\lbrace X_i\right\rbrace_{i=1}^N$, instead of encoding each element 
directly in an 8-bit floating-point format, $X$ is stored using $N$ 8-bit 
floating-point values $\left\lbrace Y_i\right\rbrace_{i=1}^N$ and two extra 
factors $\alpha$ and $\beta$ that account for statistical information about 
$X$ and capture its dynamic range.
This tensor format is called S2FP8 and its use in the training procedure 
(for the forward \& backward passes and the gradient update computations) 
is summarized in Figure~\ref{fig:fp8training_3}. 

Tests on the effectiveness of this approach (FP32 vs S2FP8) are performed on 
residual networks of varying depths on the CIFAR10 and ImageNet datasets, the 
Transformer network on an English-Vietnamese translation dataset and neural 
collaborative filtering network architecture. The authors of~\cite{cambier2020shifted}
state that the extra hardware complexity required to handle the conversion 
operations and the management of the $\alpha, \beta$ parameters at each layer
is small.

\subsection{Low precision training algorithm design}
Section~\ref{sec:mpt} reviewed how mixed-precision computation can be 
used to speed up neural network training algorithm execution, with minimal or no loss to the final test accuracy for the resulting model. Such methods
are attractive because they do not require any changes to the problem's
hyperparameters (such as learning rate scheduling), making them potentially
easy to use (for instance the use of mixed precision training with NVIDIA
GPUs is straightforward with the use of their Automatic Mixed Precision (AMP)
support for major deep learning frameworks).

An orthogonal and complementary direction is the development of learning algorithms tailored for low precision computation.  One such approach 
is MuPPET~\cite{rajagopal2020multi}, which advocates for an automatic~\emph{intra} 
epoch numerical precision switch of training quantization and computation levels. It proposes a metric that estimates how much information each new training step obtains for a given quantization level, by quantifying the diversity of computed gradients across epochs. This allows for a heuristic 
runtime policy that progressively increases the working precision/format such 
that the final test accuracy is comparable to that of baseline \texttt{float32} 
training. The approach is designed to take  advantage of the myriad of numerical 
precisions that have started to appear in modern hardware (\emph{e.g.,} 4 and 8-bit 
integer computations and 16-bit floating-point formats). For each iteration/minibatch 
and a working fixed-point precision $q$, a block floating-point training scheme 
(similar to~\cite{courbariaux2014training,
koster2017flexpoint,das2018mixed,drumond2018training})
with both values and scale factors stored as $q$-bit integers and stochastic rounding for quantization is used. Similar to most other mixed-precision approaches, 
a \texttt{float32} master copy of the weights is always kept in memory and updated at each iteration with the low precision loss function gradients computed most recently. To test this approach, five levels of precision (8-, 12-, 14- and 16-bit 
fixed-point formats and ultimately \texttt{float32}) were used in~\cite{rajagopal2020multi} 
for training AlexNet, ResNet18/20 and GoogLeNet networks with the CIFAR-10/100 
and ImageNet datasets on an NVIDIA RTX 2080 Ti GPU. A comparison with baseline 
\texttt{float32} training shows a $1.25-1.32\times$ speedup for MuPPET, whereas 
with respect to~\cite{micikevicius2017mixed}, it achieves a $1.23\times$ speedup 
for AlexNet and comparable performance for ResNet18 and GoogLeNet.

Following~\cite{de2018high} ``there is always a tradeoff with standard training algorithms:
as the number of bits is decreased, noise that limits statistical accuracy is increased''.
To limit the loss in statistical accuracy when doing low precision training, they propose 
HALP (High Accuracy Low Precision), a low precision variant of stochastic gradient descent
which uses low precision for most of the time in its innermost loop, while infrequently
recentering the weight parameters with higher precision in an outer loop to counteract the 
noise effect of low precision quantization. The idea of the algorithm is based on the 
Stochastic Variance Reduced Gradient (SVRG) approach, introduced in~\cite{johnson2013accelerating}, 
and a~\emph{bit centering} representation, where each number is represented as
the sum of a high precision offset term, modified only infrequently, and a low precision \emph{delta} term, which is modified at each inner iteration.

For strongly convex problems, the authors show that the HALP approach can produce 
arbitrarily accurate solutions retaining the same linear asymptotic convergence rate as 
SVRG in full precision. On non-convex problems (namely CNN and LSTM neural network training),
HALP (with a 16-bit low precision format and 32-bit high precision one) is empirically shown
to improve on low precision variants of SGD and SVRG and equals or outperforms full precision 
SVRG and SGD. It can also be used to effectively fine-tune low precision trained results
as well, as the authors show on a ResNet18 model, closely matching the result obtained
from de facto SGD training in full precision. On ImageNet, such variance-reduced mixed-precision training algorithms can obtain state-of-the-art timing results~\cite{jia2018highly}.

A simpler approach for a low precision training algorithm is SWALP (Stochastic
Weight Averaging in Low Precision Training)~\cite{yang2019swalp}. It is based on the recent 
\emph{Stochastic Weight Averaging} (SWA) method~\cite{izmailov2018averaging}. SWA was introduced as an SGD variant that shows improved generality in deep learning training.
    Low precision training on the other hand produces extra quantization noise and generally tends to underperform when the learning rate is low. Averaging 
    weights that have been rounded both down and up during quantization can potentially 
    reduce quantization effects and is the reason why the authors of~\cite{yang2019swalp} 
    propose that SWA can be beneficial for low precision training. The SWALP approach 
    consists of quantizing in low precision all numbers during training, including the 
    gradient accumulator (and potentially the velocity vector for momentum-based approaches).
    On a theoretical level, the authors can show that SWALP can converge to an optimal solution for quadratic objectives and a smaller noise ball than low precision 
    SGD for strongly convex objective functions. Empirically, for nonconvex objectives, 
    an 8-bit SWALP approach (with an 8-bit block floating-point format with 8-bit shared exponents) 
    can match full precision SGD baselines in DNN training tasks such as for VGG-16 and 
    Preactivation ResNet-164 on CIFAR-10/100 datasets.

\section{Support for approximation in DNN Accelerators}
\label{sec:accelerators}

DNN models can be executed in different environments, ranging from high-power data center servers to low-power edge devices. Within this large space, there is an even wider one representing the different backends that can be used. Backends are differentiated in terms of both software and hardware. The solutions vary from general-purpose frameworks and computing units to application-specific frameworks and computing units.

Like many applications, DNNs were initially executed on latency-oriented CPUs, but ever since the start of the 2010s, there has been a major shift towards parallel hardware. Examples include GPUs for performance-oriented scenarios and microcontrollers for low-power devices. Still, due to their static and general-purpose data path, General-Purpose Processors (GPPs) are not able to efficiently process DNNs in all application scenarios, motivating the need for dedicated hardware accelerators.

The first proposed hardware accelerators were ASICs~\cite{Chen2014DaDianNaoAM,Chen2016EyerissAS,Jouppi2017TPU}, and they achieved orders of magnitude improvements in energy efficiency compared to GPPs. This gain nevertheless comes at the expense of flexibility, with the design cost being very high. FPGAs, on the other hand, provide a good balance between flexibility, design cost, and performance~\cite{Guo2016AngelEye,Reddy2018DLAUAS}.

\begin{figure}[ht]
\includegraphics[width=0.5\textwidth]{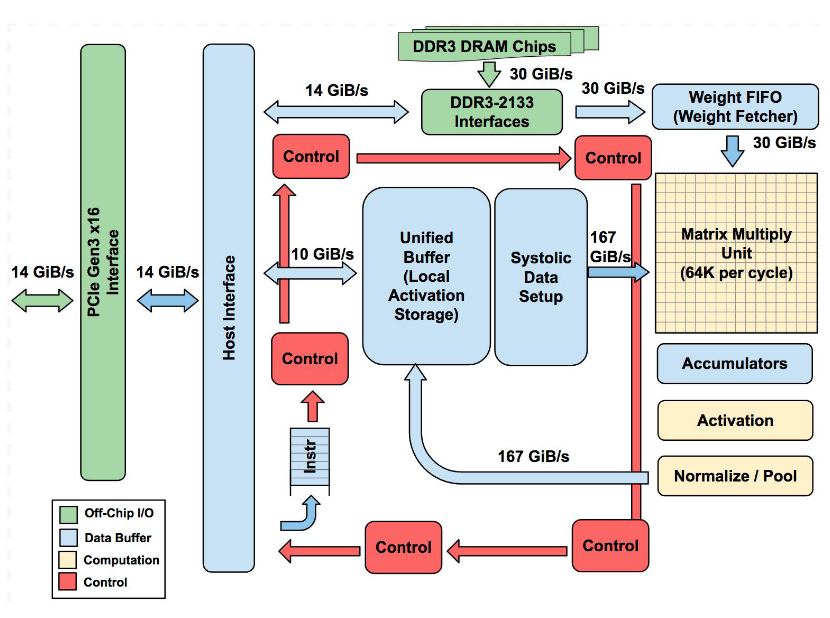}
\includegraphics[width=0.5\textwidth]{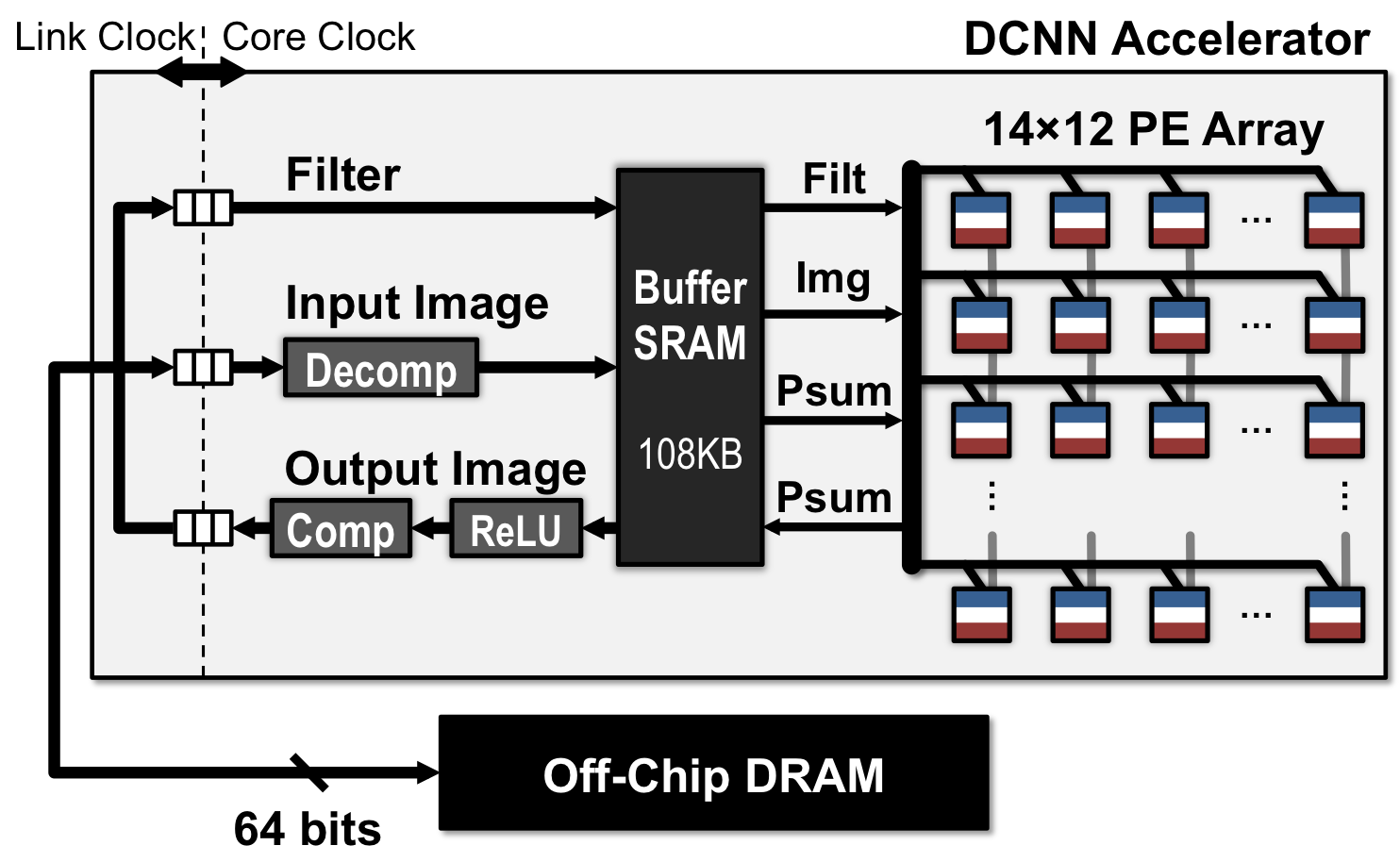}
\caption{Comparing the systolic arrays based architecture of the ASIC Google TPU~\cite{Jouppi2017TPU} (top) and the Processing Element (PE) of the FPGA grid-based Eyeriss~\cite{Chen2016EyerissAS} architecture (bottom).}
\label{fig:systolicandpe}       
\end{figure}

Independently of the target (ASIC or FPGA), hardware accelerators adopt the same strategy of maximizing data reuse, an element that has been extensively studied by Chen \& al.~in~\cite{Chen2016EyerissAS}. The main architectures adopted by re-configurable accelerators such as FPGAs is a dedicated grid of Processing Element (PE)~\cite{Chen2016EyerissAS}, while the main architecture adopted by ASICs are based on more generic systolic arrays~\cite{Jouppi2017TPU}. This is mainly because a systolic array is more flexible once designed and can efficiently process matrix products, while a PE array requires tuning some parameters for efficiently executing a DNN (like the number of PEs and the size of the memory bus), making them more suitable for re-configurable accelerators. A schematic view of the two approaches is given in Figure~\ref{fig:systolicandpe}.

Since DeepCompression~\cite{han2015deepcompress} proved that approximation techniques can significantly improve DNN processing efficiency with very small accuracy loss, approximation for DNN acceleration has become quite popular, at the same time posing new challenges for efficient processing. For example, accelerating a sparse DNN (after application of pruning methods like those presented in Section~\ref{subsec:pruning}) requires adapting the data-flow to take advantage of the available sparsity, whereas accelerating a reduced precision DNN (after application of quantization methods as described in Section~\ref{subsec:quantization}) requires implementing dedicated operators.

\label{subsec:accelinference}
\subsection{Architectures for accelerating inference}

Almost, if not all, dedicated DNN hardware accelerators rely on reduced precision computations. This is mainly because a 32-bit floating-point is not mandatory to achieve high accuracy, and has a prohibitive computing cost. Most accelerators use 16-bit or 8-bit representations, like~\cite{Chen2016EyerissAS}. Some accelerators are dedicated to specific quantization formats, such as~\cite{Guo2018FBNA} that targets acceleration of fully-binarized DNNs, or~\cite{Kudo2018LogAccelerator} for accelerating logarithmic representations.

Whereas reduced precision acceleration-based solutions mainly require changes to the arithmetic operators, accelerating pruned DNNs with a sparse representation requires changes to the dataflow. Lu \& al.~proposed to use the combination of two structures representing the COOrdinates (COO) of the values and the values as Compressed Sparse Rows (CSR)~\cite{Lu2017SparseNN}, and developed an accelerator to take advantage of these representations. It is also possible to take advantage of Feature Map (FM) sparsity. Due to the use of ReLU activations, FMs contain a large number of zeros which can be skipped during the next layer computation. CNVlutin~\cite{Albericio2016Cnvlutin} explores this dynamic sparsity. It is also possible to accelerate structured sparse DNNs with a dedicated dataflow like in~\cite{Zhu2020HWAStructuredSparseCNN}.

DNNs with shared weights can also benefit from a dedicated dataflow. This was studied in~\cite{Han2016EIE}, which targets DNNs compressed using the DeepCompression~\cite{han2015deepcompress} three-step method. It introduces an efficient
implementation of the sparse matrix-vector multiplications with weight sharing that are
central to the approach from~\cite{han2015deepcompress}.

\subsection{Architectures for accelerating training}
\label{subsec:acceltraining}

Accelerating DNN algorithms on hardware targets such as FPGAs faces many challenges, including limited on-chip memory, external memory bandwidth, and computational resources. Compared to the design of inference accelerators, on-chip training is a less studied topic, but it is feasible~\cite{tao2020challenges}.

An example is~\cite{fox2019training}, which targets training acceleration for embedded Xilinx Zynq All Programmable System on Chip (APSoC) devices. It essentially implements a version of the method introduced in~\cite{yang2019swalp} with predominantly 8-bit integer arithmetic. The Arm-based processor on the device is used for 32-bit floating-point weight updates, whereas the FPGA logic evaluates all the 8-bit integer matrix multiplications needed during the backpropagation computation path. The overall hardware platform is configured using a software-based High-Level Synthesis (HLS) flow with Xilinx tools. On the Intel side of things,~\cite{venkataramanaiah2019automatic} has proposed a Register-Transfer Level (RTL) compiler that performs SGD-based training on Intel FPGAs for various CNNs with 16-bit fixed-point arithmetic.

\section{Perspectives}\label{sec:perspectives}

Due to the rapid evolution of the field of deep learning, it is difficult to give an accurate prediction of how to approximate computing techniques that will impact DL acceleration in the future. This section presents an overview of three different research directions that figure to grow in importance in the years to come.

\subsection{Approximation for attention-based architectures}
While the focus of the previous sections is mostly directed at CNN-based models, in recent years alternative structures such as Transformer attention
architectures~\cite{vaswani2017attention} have led to state-of-the-art accuracy results in NLP-based tasks (\emph{e.g.,} language modeling). Subsequent models, like BERT~\cite{devlin2018bert}, RoBERTa~\cite{liu2019roberta} and GPT~\cite{brown2020language}, although impressive, have a large memory footprint, increased latency, and power consumption that are prohibitive for efficient deployment on embedded edge devices and even on data centers. Due to their expressive power, Transformer-based models are also beginning to be adapted for other tasks, such as computer vision applications~\cite{carion2020end,strudel2021segmenter}.

Their increasing usage is driving interest for efficient approximation methods that specifically target Transformer models. While work in this direction is still in its early stages, there are already some approaches based on quantization~\cite{zadeh2020gobo,kim2021bert}, knowledge distillation~\cite{sanh2019distilbert,jin2021kdlsq} and pruning~\cite{mao2021tprune,wang2020spatten}.

\subsection{Edge AI}

One area where training acceleration with reduced precision and increased energy efficiency is becoming important is incremental/lifelong learning scenarios on edge devices (\emph{e.g.,}~in autonomous driving, IoT, and robotics). Compared to a cloud-based scenario, training locally avoids transferring data back and forth between data centers and IoT devices, helping reduce communication and latency and improve privacy. 

Such on-chip training is feasible~\cite{tao2020challenges}, but extremely challenging. The training acceleration methods described in Section~\ref{sec:training} usually cannot be applied directly to this context and alternatives need to be considered.

A training framework specifically designed for such scenarios is E$^2$-Train~\cite{wang2019e2}, 
which proposes three complementary strategies: (a) stochastic mini-batch dropping to eliminate what 
can be considered ``unnecessary costs'', (b) input-dependent selective layer update where a different subset of 
CNN layers are updated for every minibatch, and (c) predictive sign gradient descent, a variation of an extremely low precision SGD algorithm, signSGD~\cite{bernstein2018signsgd}. Besides this approach, other algorithmic \& arithmetic-level methods have started to appear~\cite{fu2020fractrain, fu2021cpt}. It is expected that this area of research will grow in importance in the years to come, with on-site learning becoming paramount in certain application domains.

\subsection{Analog in-memory computing}
The recent explosive growth in highly data-centric applications related to DL has motivated the appearance of analog in-memory computing solutions~\cite{shafiee2016isaac,chi2016prime,ankit2019puma,seb2019cim,demler2018mythic} as alternatives to traditional von Neumann computing systems. Hereby important computational tasks, such as vector-matrix multiplications, are performed in place in the memory itself by exploiting the physical attributes of the memory devices (\emph{e.g.,}~Kirchhoff’s current summation law). Besides alleviating the costs in latency and energy associated with data movement, in-memory computing also has the potential to significantly improve the computational time complexity by using large crossbar memory arrays~\cite{sebastian2020memory}. However, this comes at the expense of imprecision in the mixed-signal computations and becomes a form of approximate computing. For instance, the mapping of synaptic weights onto some of those memory devices suffers from non-ideal analog storage in the form of stochastic distribution of conductance values and temporal drifting. Accordingly, Joshi et al.~\cite{joshi2020accurate} have proposed a custom noise-injection training method to increase the robustness of the resulting network to such non-idealities and achieve a software equivalent accuracy. Given the game-changing advantages in computing efficiency of analog in-memory computing, more work is expected on this nascent field in the future.


%
%

\section{Conclusion}
\label{sec:conclusions}

In this chapter, a comprehensive survey of approximation techniques applied to Deep Learning is provided. These techniques target various improvements, some geared towards the training of DNN models, others that focus on DNN inference. Depending on the objective, various methods can be applied, whether for the improvement (reduction) of memory requirements by using compression techniques or for the reduction of the computational workload by using acceleration techniques.

Such a wide range of approximation techniques involves various implementation changes, ideally resulting in a backend adaptation that maximizes the expected performance improvement. These adaptations can be implemented at the software level using dedicated frameworks and/or at the hardware level in dedicated accelerators.

To compare the various methods available, it is desirable to use the same input DNN and workload, but this is not always feasible. This is mostly due to the large range of DNN topologies that have appeared over the years: while some methods can be applied almost automatically to multiple topologies, some require manual tuning as the size of the search space increases exponentially with the DNN size. There is also a wide range of workloads, from small ``toy'' datasets to more recent and challenging large-scale datasets. Some methods cannot perform equally well in both contexts. The difference in backend compatibility with the various approximation methods also regularly involves manual tuning steps, which are hard to reproduce and compare to other backends.

Most of the recent methods give very promising results and pave the way for further research, by proving that approximations can be applied at various levels, from the topology of the DNN, to the data value and type, and including the backends-DNN codesign (hardware or software).

\section{Acknowledgments}
\label{sec:acknowledgments}

This work has been funded by the French National Research Agency (ANR) through the AdequatedDL research project (ANR-18-CE23-0012).

\bibliographystyle{alpha}
\bibliography{references}

\end{document}